\documentclass[12pt,preprint]{aastex}

\usepackage{color}

\slugcomment{}

\shorttitle{Infall in the Circumbinary Disk around L1551 NE}
\shortauthors{Takakuwa et al.}

\begin{document}


\title{Angular Momentum Exchange by Gravitational Torques and Infall\\
in the Circumbinary Disk of the Protostellar System L1551 NE}

\author{Shigehisa Takakuwa\altaffilmark{1}, Masao Saito\altaffilmark{2},
Kazuya Saigo\altaffilmark{3}, Tomoaki Matsumoto\altaffilmark{4},
Jeremy Lim\altaffilmark{5}, Tomoyuki Hanawa\altaffilmark{6},
\& Paul T. P. Ho\altaffilmark{1}}
\altaffiltext{1}{Academia Sinica Institute of Astronomy and Astrophysics, P.O. Box 23-141, Taipei 10617, Taiwan;
takakuwa@asiaa.sinica.edu.tw}
\altaffiltext{2}{Joint ALMA Observatory, Ave. Alonso de Cordova 3107, Vitacura, Santiago, Chile}
\altaffiltext{3}{ALMA Project Office, National Astronomical Observatory of Japan, Osawa 2-21-1,
Mitaka, Tokyo 181-8588, Japan}
\altaffiltext{4}{Faculty of Humanity and Environment, Hosei University, Chiyoda-ku, Tokyo 102-8160}
\altaffiltext{5}{Department of Physics, University of Hong Kong, Pokfulam Road, Hong Kong}
\altaffiltext{6}{Center for Frontier Science, Chiba University, Inage-ku, Chiba 263-8522, Japan}

\begin{abstract}
We report the ALMA observation of the Class I binary protostellar system L1551 NE
in the 0.9-mm continuum, C$^{18}$O (3-2), and $^{13}$CO (3-2) lines at a $\sim$1.6 times higher resolution
and a $\sim$6 times higher sensitivity than those of our previous SMA observations, which
revealed a $r$ $\sim$300 AU-scale circumbinary disk in Keplerian rotation.
The 0.9-mm continuum shows two opposing $U$-shaped brightenings in
the circumbinary disk, and exhibits a depression between the circumbinary disk and the circumstellar disk of the
primary protostar. The molecular lines trace non-axisymmetric deviations from Keplerian rotation
in the circumbinary disk at higher velocities relative to the systemic velocity, where our previous SMA
observations could not detect the lines.
In addition, we detect inward motion
along the minor axis of the circumbinary disk. To explain the newly-observed features, we performed a numerical
simulation of gas orbits in a Roche potential tailored to the inferred properties of L1551 NE. The observed $U$-shaped
dust features coincide with locations where gravitational torques from the central binary system are predicted to impart
angular momentum to the circumbinary disk, producing shocks and hence density enhancements seen as a pair of spiral
arms. The observed inward gas motion coincides with locations where angular momentum is predicted to be lowered by
the gravitational torques. The good agreement between our observation and model indicates that gravitational torques
from the binary stars constitute the primary driver for exchanging angular momentum so as to permit infall through
the circumbinary disk of L1551 NE.
\end{abstract}

\section{Introduction}

One of the outstanding questions in astrophysics is how angular momentum is exchanged
in what would otherwise be centrifugally-supported disks so as to permit infall through the disk
and accretion onto the central object ($e.g.$, star or stellar remnant, or a galactic central
super-massive black hole).
Gravitational instabilities \cite{kra10,vor10,vor11} and magneto-rotational instabilities
\cite{bal03,ma11a,ma11b}
are often considered to transfer angular momentum and permit infall in the disks.
Interferometric observations at millimeter and submillimeter wavelengths
have revealed an increasing number of circumstellar disks around
Class I protostars\cite{bri07,lom08,jor09,yen13,har14,yen14}.
Recent observations have also found
disks around Class 0 protostars \cite{tob12,mur13,lin14},
as well as circumbinary disks around both
components of a binary protostellar system \cite{tak04,tak12,tob13}.
Mechanisms must operate in these disks so as to permit infall through the disk and accretion
onto the central protostar or protostellar system.

Here, we address the forces that may be responsible for driving the exchange of
angular momentum in circumbinary disks so as to permit infall through these disks
and hence growth of the binary system.
The formation of binary systems is the primary mode of star formation at solar masses
\cite{mat94,mat00,che13}.
One commonly examined mode in which such systems may form is through the fragmentation
of a massive, rotating, disk-like structure due to the gravitational instability
\cite{mat03,nak03,mac08,kra10,vor10,vor11,zhu11}
(see the argument by Maury et al. 2010 too).
In this scenario, the resulting protostellar system comprises two (or more) protostars,
each exhibiting a circumstellar disk, surrounded by a rotationally-supported circumbinary disk
\cite{art96,ba97b,bat00,gun02,och05,han10}.
Gravitational torques from the binary protostars exert a non-axisymmetric force on the circumbinary disk,
creating most visibly a characteristic two-arm spiral (in the case of binary systems) where the local
angular momentum in the circumbinary disk is elevated \citep[e.g.,][]{ba97b}.
Infall occurs away from the spiral arms and through the outer Lagrangian points of the binary system
where the local angular momentum in the circumbinary disk is lowered.
Until now, however, observations have not had sufficient angular resolution and sensitivity to
detect such features in circumbinary disks.
Here, we report our observation of L1551 NE
using the Atacama Large Millimeter/submillimeter Array (ALMA) designed to test
the abovementioned model predictions for its circumbinary disk.

L1551 NE is a Class I protostellar system ($T_{bol}$ = 91 K,
$L_{bol}$ = 4.2 $L_{\odot}$) located in the Taurus Molecular Cloud
\citep{sai01,yok03,fro05}
at a distance of 140 pc \citep{tor09}.
Observations with the Very Large Array (VLA) reveals
two 3.6-cm radio continuum sources with a projected separation of $\sim$70 AU
at a position angle of 120$\degr$: the south-eastern source is referred to as ``Source A" and the
north-western source ``Source B" \citep{rei02}.
Near-infrared observations of L1551 NE reveal that Source A drives
collimated [Fe II] jets along the northeast to southwest direction,
and Source B is located at the origin of an extended ($\sim$2000 AU)
NIR reflection nebula with its symmetry axis at essentially the same position angle as the jets
from Source A \cite{rei00,rei02,hay09}.
Our previous observations of L1551 NE with the SubMillimeter Array (SMA)
in the 0.9-mm continuum, C$^{18}$O (3-2), and $^{13}$CO (3-2) lines
revealed a weakly extended central dust component associated with the two
protostellar components identified as their circumstellar disks, surrounded by a ring-like circumbinary disk
having an outer radius of $\sim$300 AU \cite{tak12}.
The major axis of the circumbinary disk is orthogonal to the [Fe II] jets
from Source A and the outflow cavity associated with Source B,
suggesting that the circumbinary disk is aligned with the circumstellar disks
of the binary components.
Based on the inferred alignment of both the jets and outflow cavity with respect to
the plane of the sky, the near side of the circumbinary disk is to the east and the far side to the west.
The gas motion in the circumbinary disk as measured in our SMA observations can be satisfactorily
modeled as pure (circular) Keplerian rotation, from which we derived a total mass
for the binary system of $\sim$0.8 $M_{\odot}$,
as well as an inclination of $i\sim$62$\degr$ for the circumbinary disk and a position angle for its
major axis of $\theta\sim$167$\degr$.
Based on the projected orbital separations of Source A and Source B from the
inferred kinematic center of the circumbinary disk, we derive a mass ratio for the binary protostars
of $M_B$ / $M_A$ $\sim$0.19 where $M_A$ and $M_B$, respectively, denote
the masses of Sources A and B.
Our follow-up observations of L1551 NE at lower angular resolutions in the C$^{18}$O (3-2) line
with the SMA revealed that the circumbinary disk is surrounded by an infalling envelope that
can be traced out to a radius of $\sim$500 AU \cite{tak13}.

Our observation of L1551 NE with ALMA were made during the first cycle of regular scientific
observations (Cycle 0) with this instrument, and like in our SMA observations were made 
in the 0.9-mm dust continuum, C$^{18}$O ($J$=3-2), and $^{13}$CO ($J$=3-2) lines.
At a spatial resolution that is 1.6 times higher (in beam area)
and sensitivity that is $\sim$6 times better (in brightness temperature) than those
attained in our previous SMA observations, we are able to detect for the first time internal structures and
non-axisymmetric gas motion in the circumbinary disk of L1551 NE.
To aid in interpreting the results, we performed a 3-dimensional hydrodynamic simulation
of the distribution and motion of matter around a binary protostellar system
tailored to the inferred properties of L1551 NE.
We then conducted radiative transfer calculations
to produce theoretically-predicted images that we then compared with the observed images.
In Section 2, we describe our observation and data reduction. In Section 3, we present
the results and describe the new features revealed in our ALMA observation. In Section 4,
we describe our theoretical model of L1551 NE, and compare the model to
our observation to determine whether the observed features can be reproduced solely
by gravitational torques without having to invoke magnetic fields.
In Section 6, we provide a concise summary of the results and our interpretation.
As supplementary information the velocity channel maps of the $^{13}$CO emission
are shown in Appendix A.
A more detailed technical description of how we subtracted Keplerian rotational motion
from the C$^{18}$O image cube can be found in Appendix B, and our theoretical modeling
in Appendix C.

\section{ALMA Observation and Data Reduction}

Our observation of L1551 NE was performed on 2012 November 18
with ALMA in its extended configuration during the Cycle 0 stage of scientific observations.
Table 1 summarizes the observational parameters. During the observation,
the precipitable water vapor in the atmosphere was stable
at a mean value of $\sim$0.96 mm. Excluding overheads for calibration, the
total time on source is 78 minutes.
The minimum projected $uv$ distance is 16.368 $k\lambda$
at the C$^{18}$O frequency (329.3305453 GHz),
implying that we are able to recover virtually all the flux of the circumbinary disk out to
its outermost radial extent of $\sim$300 AU \cite{wil94}.
On the other hand, we are largely insensitive to its surrounding envelope,
such that only 10$\%$ of the peak flux is recoverable for a Gaussian
emission distribution with a FWHM of 10$\arcsec$ ($\sim$1400 AU).

The ALMA correlator was configured in the Frequency
Division Mode (FDM) to provide four independent spectral windows,
each having a bandwidth of 468.75 MHz.
The individual spectral windows are divided into 3840 channels,
with each channel having a width of 122.07 kHz.
Hanning smoothing was applied to the spectral channels,
resulting in a frequency resolution of 244.14 kHz and hence
a velocity resolution of 0.22 km s$^{-1}$ at the
C$^{18}$O ($J$=3-2) frequency.
With one spectral window dedicated to the C$^{18}$O (3-2) line,
the remaining three spectral windows were assigned to the
$^{13}$CO ($J$=3-2; 330.587965 GHz),
HC$^{18}$O$^{+}$ ($J$=4-3; 340.63070 GHz),
SO ($J_N$=7$_8$-6$_7$; 340.71416 GHz), and CS ($J$=7-6; 342.882857 GHz) lines.
Channels in all four spectral windows devoid of line emission
were used to create the continuum image, which has
a central frequency of 335.85 GHz (= 0.893 mm) and a total bandwidth of 1.79 GHz.
In this paper, we focus on the results of the continuum and C$^{18}$O (3-2) line
only because they trace the structure and kinematics of
the circumbinary disk most clearly (we plan to report the full results in the other lines in a separate paper).
As described below, the newly-found velocity features
in C$^{18}$O are also seen in the other lines at
velocities sufficiently far away from the systemic velocity.
Closer to the systemic velocity, the spatial-kinematic structure in the other lines is complicated by the effects
of missing flux as well as an outflow (see Appendix A).

Calibration of the raw visibility data was performed using the
Common Astronomy Software Applications (CASA) program as implemented in a
standard reduction script for Cycle 0 data.
The absolute flux value of the amplitude calibrator, Callisto,
was derived from the CASA Butler-JPL-Horizons 2010 model,
and is accurate to within a systematic flux uncertainty of 10$\%$.
To create the image cubes in molecular lines, we applied a
Fourier transform to the calibrated data to produce an intermediate (``DIRTY'') image,
and deconvolved from this image the sidelobes of the telescope point spread function to
create the final (``CLEAN'') image. Briggs weighting having a robust parameter
of +0.5 was adopted in the C$^{18}$O and $^{13}$CO images to give
the best compromise between spatial resolution and signal-to-noise ratio.
The same procedure was adopted for the continuum image, except that we
adopted the Natural weighting, robust=+0.5 weighting, and the Uniform weighting
using the data with projected $uv$ distances larger than 80 $k\lambda$
(see Figure \ref{cont}). The latter imaging is to
best emphasize the circumstellar disks and features in the circumbinary disk,
and we confirmed that the synthesized beam is still clean with the primary sidelobe
level of $\sim$25$\%$.

\placetable{tbl-1}

\section{Results}
\subsection{0.9-mm Continuum}

\begin{figure}
\epsscale{1.0}
\plotone{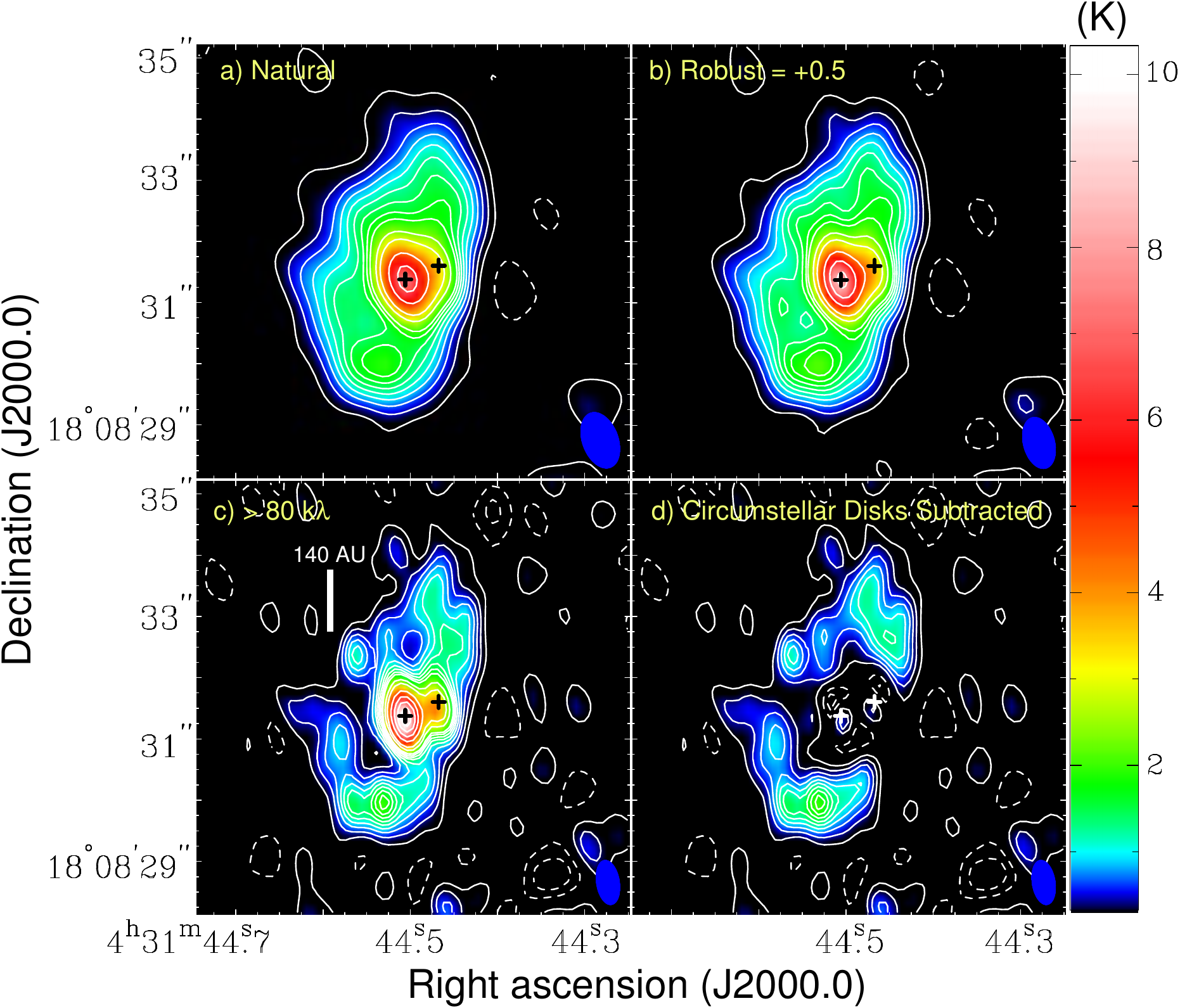}
\caption{ALMA 0.9-mm dust-continuum images of L1551 NE
with different imaging methods; a) Image with the Natural weighting,
b) Image with the robust parameter of +0.5, c) Image using the
visibility data at the $uv$ distance higher than 80$k \lambda$
with the Uniform weighting, and d) same as that in panel $c$
but after the subtraction of the central circumstellar-disk components.
Contour levels start from 0.107 K $\times$2 in steps of $\times$2 until $\times$18,
and in steps of $\times$6 until $\times$30, and then in steps of
$\times$15.
1$\sigma$ noise levels of the images in panels a, b, and c and d are
0.105 K, 0.090 K, and 0.107 K, respectively.
Filled ellipses at the bottom-right corners show the synthesized beams,
and the beam sizes are 0$\farcs$94$\times$0$\farcs$57 (P.A.=20.2$\degr$),
0$\farcs$85$\times$0$\farcs$50 (P.A.=15.2$\degr$), and
0$\farcs$72$\times$0$\farcs$36; P.A.=9.1$\degr$ for panels a, b, and c and d,
respectively.
The lower-left and upper-right crosses indicate the centroid positions of
the 2-dimensional Gaussian fittings to the dusty components associated
with Sources A and B, which we regard as the positions of Sources A and B.
\label{cont}}
\end{figure}

Figures \ref{cont}a, b, and c show the 0.9-mm continuum images of L1551 NE
at the Natural weighting, robust=+0.5 weighting, and the Uniform weighting
using the data with the projected $uv$ distances larger than 80 $k\lambda$.
These three images demonstrate the circumbinary structures at progressively
higher angular resolutions, without missing the primary emission components.
In Figures \ref{cont}a and b, a compact component
associated with Source A, with an emission extension to the location of Source B,
is seen. This component most likely traces the circumstellar disks around
each binary members.
This circumstellar-disk component is embedded in an extended emission
elongated along the northwest to southeast direction, which most likely traces
the circumbinary disk.
In the highest-resolution (0$\farcs$72$\times$0$\farcs$36; P.A.=9.1$\degr$)
0.9-mm image shown in Figure \ref{cont}c,
the central emission component seen in the lower-resolution images is
resolved into two compact components, verifying the presence of
the circumstellar disks.
Gaussian fits to these two
components demonstrate that they are barely if at all resolved.
The south-eastern component has a flux density of 0.35 Jy and a centroid location of
(04$^{h}$31$^{m}$44$\fs$51, 18$\degr$08$\arcmin$31$\farcs$4),
and the north-western component a flux density of 0.17 Jy and a centroid location of
(04$^{h}$31$^{m}$44$\fs$47, 18$\degr$08$\arcmin$31$\farcs$6).
The positions of these components are slightly ($\sim$0$\farcs$3) south
of the positions of Sources A and B derived from previous observations
at 3.6 cm that trace free-free emission (outflows) from the individual protostars \cite{rei02},
but consistent within $\lesssim$0$\farcs$1 with those measured in our recent
observations with the JVLA at 7 mm \cite{lim14}. The slight positional shift over the last decade is
consistent with the absolute proper motion of the L1551 region found by
Jones \& Herbig (1979) and Rodr\'iguez et al. (2003)
($\mu_{\alpha}$ = 0$\farcs$012 yr$^{-1}$, $\mu_{\delta}$ = -0$\farcs$023 yr$^{-1}$).
Hereafter, we regard the centroid positions of these components as derived from their
Gaussian fits as the positions of the individual protostars.
In addition to the circumstellar disks, the highest-resolution image reveals,
for the first time, brightenings in the circumbinary disk.
The most prominent internal brightenings comprise
the $U$-shaped feature to the south of the binary system as well as protrusions
to the north of Source B and north-east of Source A.
A depression or gap in the dust emission also can be seen
between the northern and southern parts of the circumbinary disk and
the circumstellar disk of Source A.

To better separate the circumstellar disks from
the circumbinary disk, we subtracted two Gaussian components
with parameters derived from the fit to the two compact dust components.
The resultant image is shown in Figure \ref{cont}d, where the
southern $U$-shaped feature remains evident.
The protrusions to the north of Sources A and B can now be seen to comprise another
$U$-shaped feature, but now located to the north of the binary system.
As we will show below, the northern and southern $U$-shaped features correspond to
spiral arms (Arms A and B, respectively) produced by the gravitational torques
from the binary stars.

We estimate the masses of the individual circumstellar disks and circumbinary disk ($\equiv M_{d}$)
from their individual continuum fluxes ($\equiv S_{\nu}$) using the relationship
\begin{equation}
M_{d}=\frac{S_{\nu}d^2}{\kappa_{\nu} B_{\nu}(T_d)},
\end{equation}
where $\nu$ is the frequency, $d$ the distance, $B_{\nu}(T_d)$ the Planck function
for dust at a temperature $T_{d}$, and $\kappa_{\nu}$ the dust opacity per unit gas + dust mass
on the assumption of a gas-to-dust mass ratio of 100.
We adopt the relation $\kappa_{\nu}$ = $\kappa_{\nu_{0}}$($\nu$/$\nu_{0}$)$^{\beta}$,
where $\beta$ denotes the dust-opacity index and
$\kappa_{250~\mu m}$=0.1 cm$^{2}$ g$^{-1}$ \cite{hil83},
as is widely adopted in the literature for circumstellar disks around low-mass protostars.
Recent multi-frequency observations
of circumstellar disks in dust emission show that $\beta$ is in the range
of $\sim$0 -- 1 \cite{gui11,chi12}, and so here we adopt $\beta$=0.5.
The dust mass opacity at 0.9 mm is then calculated to be
$\kappa_{0.9mm}$ = 0.053 cm$^{2}$ g$^{-1}$. This value is a factor 3 higher
than that of Ossenkopf \& Henning (1994) for grains
with thin ice mantles coagulated at a density of 10$^{6}$ cm$^{-3}$
($\kappa_{0.9mm}$ = 0.018 cm$^{2}$ g$^{-1}$). Thus, adopting the dust mass
opacity by Ossenkopf \& Henning (1994) provides a factor 3 higher masses
of the circumstellar and circumbinary disks.
Regarding the dust temperature, modeling of the spectral energy distribution (SED)
toward L1551 NE as measured in
single-dish observations at wavelengths from 12 $\mu$m to 2 mm
derived $T_{d}$ $\sim$42 K \cite{bar93,mor94}.
This temperature is, however, likely to be biased by the optically-thick
mid-infrared emission ($\sim$12 -- 100$\micron$), and thus the temperature
is more likely the surface temperature of the disk.
The dust temperature in the disk midplane of Class I sources can be as low as $\sim$10 K
\cite{naz03,whi03}. With $T_{d}$ = 10 -- 42 K and $\kappa_{0.9mm}$ = 0.053 cm$^{2}$ g$^{-1}$, 
a mass for the circumstellar disk of Source A is calculated to be
$\sim$0.005 - 0.044 $M_{\odot}$, that of Source B a mass of $\sim$0.003 - 0.022 $M_{\odot}$.
The flux density of the circumbinary disk in the 0.9 mm continuum as computed
from Figure \ref{cont}d is 0.47 Jy, corresponding to a mass for the circumbinary disk of
$\sim$0.007 - 0.059 $M_{\odot}$.
The individual as well as combined disk masses are therefore much smaller than the
inferred total binary mass of 0.8 $M_{\odot}$.
In our theoretical model for the distribution and dynamics of matter around L1551 NE,
we therefore ignore the self-gravity of the circumstellar disks and circumbinary disk,
and only consider the gravitational field of the binary protostars.

\placefigure{cont}

\subsection{Molecular Line}
\subsubsection{Integrated Intensity Map}

\begin{figure}
\epsscale{1.0}
\plotone{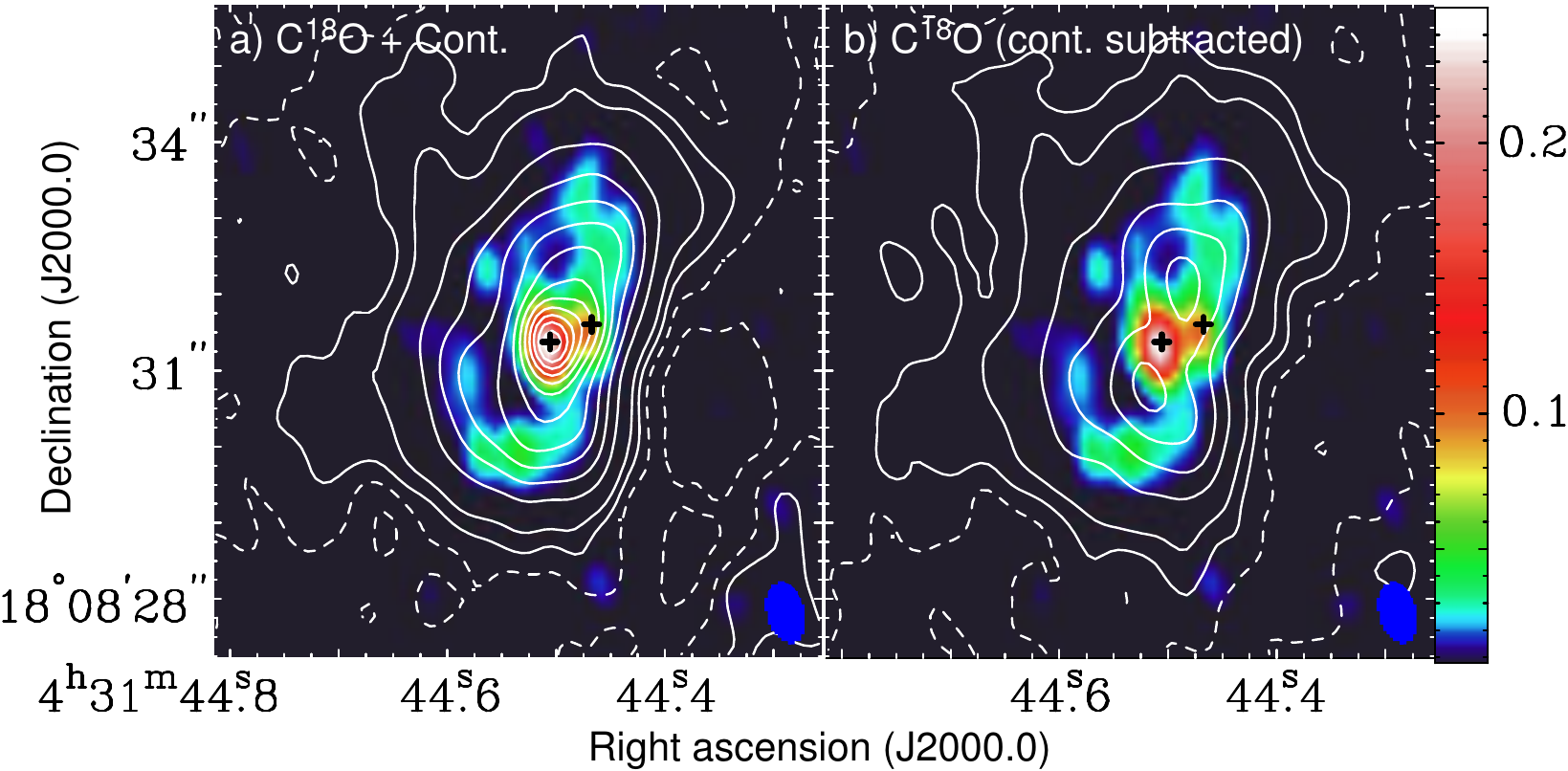}
\caption{a) ALMA image of the C$^{18}$O ($J$=3-2) plus continuum emission of L1551 NE
integrated over the velocity range from $V_{\rm LSR}$=2.67 km s$^{-1}$ to 10.56 km s$^{-1}$,
superposed on the 0.9-mm dust-continuum image of L1551 NE
(colors; same as in Figure \ref{cont}).
Contour levels are 3$\sigma$, 10$\sigma$, 30$\sigma$, and then in steps of 20$\sigma$
(1$\sigma$ = 14.8 mJy beam$^{-1}$ km s$^{-1}$).
Crosses denote the positions of the protostellar binary, and
a filled ellipse at the bottom-right corner the synthesized beam of the C$^{18}$O plus continuum image
(0$\farcs$82$\times$0$\farcs$49; P.A.=15.9$\degr$).
b) ALMA image of the C$^{18}$O ($J$=3-2) emission of L1551 NE after the subtraction of
the continuum level (contours), superposed on the 0.9-mm dust-continuum image (colors).
Contour levels and symbols are the same as those in panel a).
\label{c18omom0s}}
\end{figure}

Figure \ref{c18omom0s} shows the C$^{18}$O integrated intensity map
at an angular resolution of 0$\farcs$82$\times$0$\farcs$49 (P.A.=15.9$\degr$) (contours)
superposed on the 0.9-mm dust-continuum map (colors; same as that in Figure \ref{cont}d)
of L1551 NE. The C$^{18}$O map in Figure \ref{c18omom0s}a includes an unknown contribution from dust,
whereas that in Figure \ref{c18omom0s}b has a constant level subtracted from every channel
corresponding to the dust continuum emission as measured outside the velocity range
spanned by the C$^{18}$O line.
The C$^{18}$O map before continuum subtraction shows emission from
the circumstellar disks, whereas that after subtracting a constant continuum component
shows little if any such emission.
In the circumstellar disks, both the dust and gas opacities are appreciable, resulting in partial absorption
of the line emission by the dust and partial absorption of the continuum emission by the gas.
In this situation, it is not straightforward to separate the line and continuum emission.
The same complexity is seen, for example, in the circumstellar disk around
the T-Tauri star HD 142527 \cite{fuk13}.
Beyond the circumstellar disks where the dust opacity is significantly lower,
the intensity distribution in C$^{18}$O is similar irrespective of whether a constant continuum
component is subtracted or not.
The circumbinary disk can be traced out to a radius of $\sim$300 AU in C$^{18}$O,
further out than in the dust continuum.
By contrast with the dust continuum map, the C$^{18}$O map shows no depression
between the circumbinary disk and the circumstellar disk of Source A, but instead
increases inwards in intensity towards both circumstellar disks.
The observed behavior suggests that C$^{18}$O is more optically thick
than dust, and that the region between the circumbinary disk and the circumstellar disk of Source A
is not devoid of matter.

\placefigure{c18omom0s}

\subsubsection{Velocity Structure}

\begin{figure}
\epsscale{0.8}
\plotone{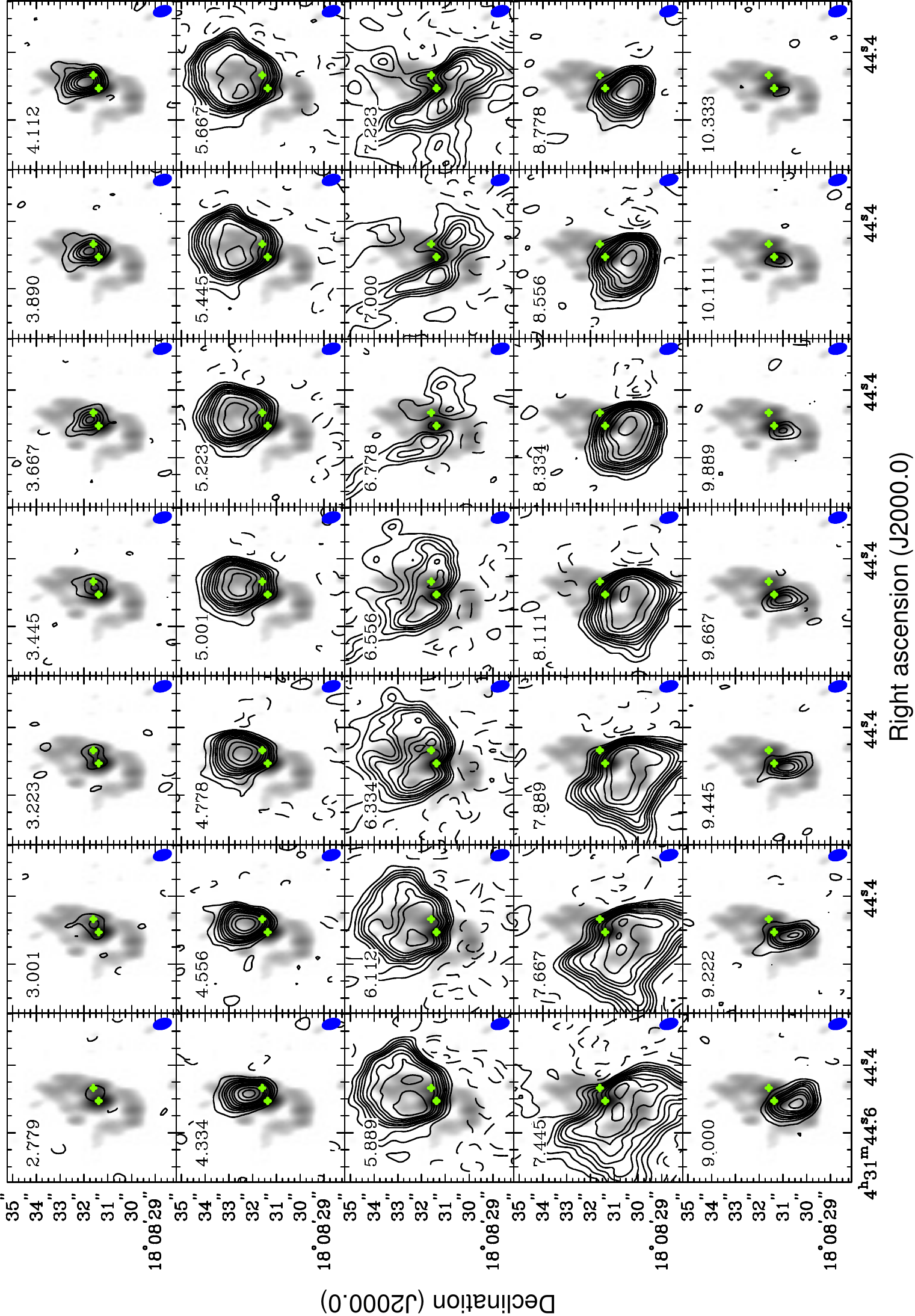}
\caption{Observed velocity channel maps of the C$^{18}$O ($J$=3-2) emission
(contours) superposed on the 0.9-mm dust-continuum emission (grey) in L1551 NE,
taken with ALMA. Numbers at the top-left corners denote the LSR velocities.
Contour levels start from 3$\sigma$ in steps of 3$\sigma$ until 15$\sigma$,
and 20$\sigma$, 25$\sigma$, 30$\sigma$, 40$\sigma$, 50$\sigma$, and then
in steps of 20$\sigma$ (1$\sigma$ = 11.2 mJy).
Crosses show the positions of the protostellar binary, and
filled ellipses at the bottom-right corners the synthesized beam
(0$\farcs$82$\times$0$\farcs$49; P.A.=15.9$\degr$).
\label{c18och}}
\end{figure}

Figure \ref{c18och} shows the continuum-subtracted velocity channel maps of the C$^{18}$O (3-2) line
(contours) superposed on the 0.9-mm dust-continuum image (grey scale).
In our previous observations of L1551 NE with the SMA,
we established that C$^{18}$O has a systemic velocity of $V_{\rm LSR}$=6.9 km s$^{-1}$,
and detected this line over the velocity range from $\sim$4.3 km s$^{-1}$ to $\sim$9.1 km s$^{-1}$
\cite{tak12,tak13}.
In our observation here at a higher sensitivity using ALMA, C$^{18}$O can be detected
to higher blueshifted velocities of 2.8 -- 4.3 km s$^{-1}$ and redshifted velocities of
9.0 -- 10.3 km s$^{-1}$. As can be seen in Figure \ref{c18och}, the C$^{18}$O emission
peaks between Sources A and B at blueshifted velocities
of 2.8 -- 3.5 km s$^{-1}$, and to the southeast of Source A at
redshifted velocities of 9.9 -- 10.3 km s$^{-1}$. As we will explain later,
these high-velocity components likely trace gas in the Roche lobes of the two protostars.
At blueshifted velocities of 3.7 -- 4.8 km s$^{-1}$
the C$^{18}$O emission extends northeast from the midpoint between Sources A and B, and
at redshifted velocities of 8.3 -- 9.7 km s$^{-1}$ to an area spanning from south to southwest of both sources.
At velocities of 5.0 -- 6.1 km s$^{-1}$ and 7.4 -- 8.1 km s$^{-1}$,
the C$^{18}$O emission originates predominantly from the northern and southern parts,
respectively, of the circumbinary disk.
These low-velocity blueshifted and redshifted emissions trace
the outer northern and southern parts of the circumbinary disk seen in the 0.9-mm
continuum emission, but extends further out that in the continuum.
Close to the systemic velocity (6.3 -- 7.2 km s$^{-1}$), the C$^{18}$O emission
exhibits a ``butterfly'' pattern (particularly at 6.8 and 7.0 km s$^{-1}$),
a characteristic signature of a Keplerian disk \cite{sim00}.
The channel maps at velocities relatively close to systemic (5.0-8.1 km s$^{-1}$) trace
the same regions as previously seen in our SMA observations,
where the spatial-kinematic structure of the circumbinary gas out to a radius of
$\sim$300 AU was satisfactorily modeled as a Keplerian disk.
On the other hand, as we shall show, the emission extending to higher blueshifted and
redshifted velocities detected here for the first time trace deviations from Keplerian rotation.

\begin{figure}
\epsscale{1.0}
\plotone{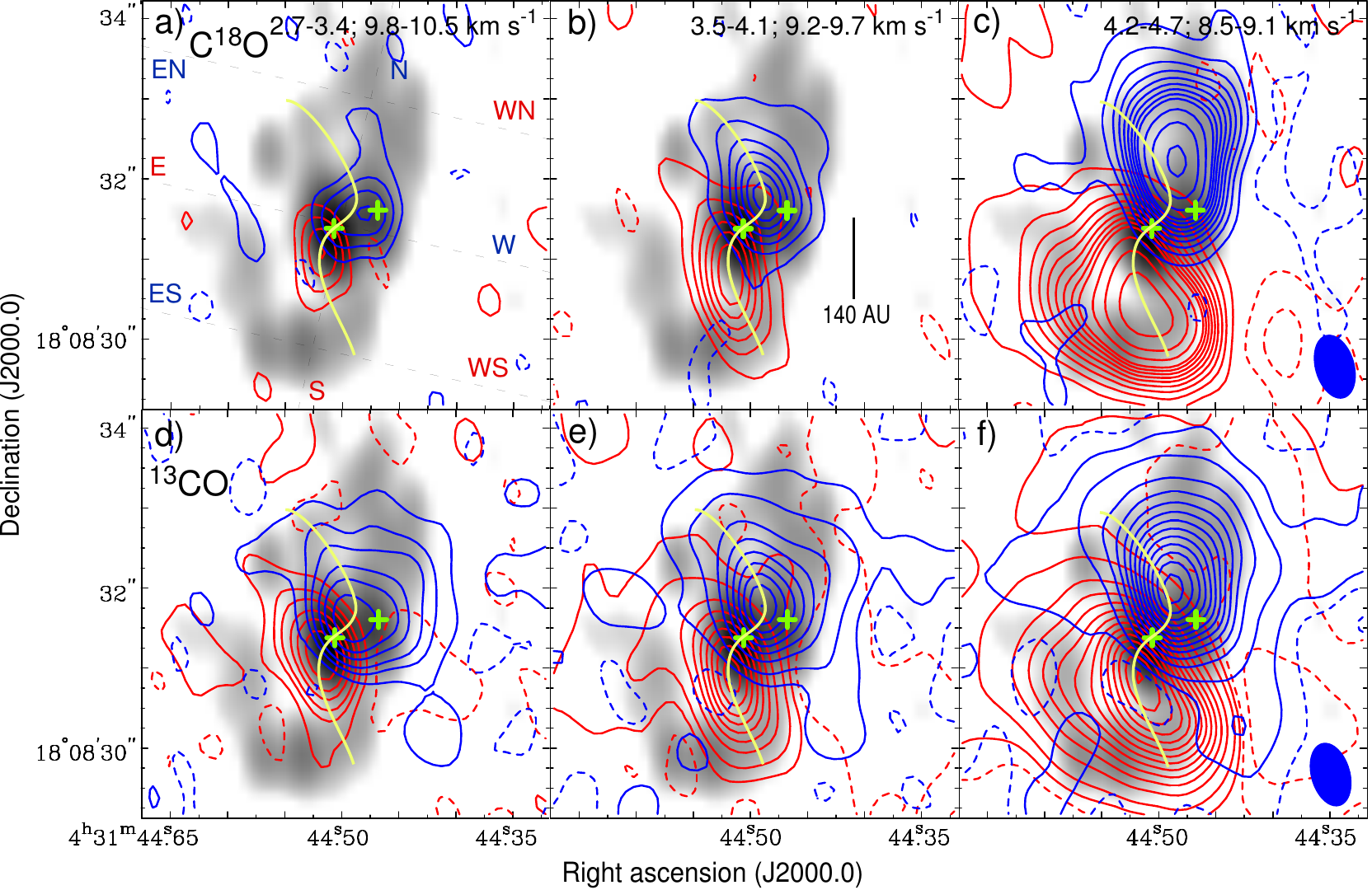}
\caption{Maps of high-velocity blueshifted (blue contours) and redshifted (red)
C$^{18}$O ($J$=3-2) (upper panels) and $^{13}$CO ($J$=3-2) emission (lower)
in the velocity regions as labeled, superposed on the 0.9-mm
dust-continuum image (grey scale) in L1551 NE.
In panel a) contour levels are in steps of 3$\sigma$ (1$\sigma$ = 6.0 mJy beam$^{-1}$).
In panels b) and c) contour levels start from
3$\sigma$ in steps of 5$\sigma$ until 43$\sigma$, and then 55$\sigma$, 70$\sigma$, and
90$\sigma$ (1$\sigma$ = 6.5 mJy beam$^{-1}$).
In panel d) contour levels are 3$\sigma$, 9$\sigma$, 15$\sigma$, 25$\sigma$,
35$\sigma$, and 50$\sigma$ (1$\sigma$ = 4.15 mJy beam$^{-1}$).
In panels e) and f) contour levels are 3$\sigma$, 13$\sigma$, and
then in steps 20$\sigma$ (1$\sigma$ = 4.5 mJy beam$^{-1}$).
Crosses show the positions of the protostellar binary, and
a filled ellipse at the bottom-right corner in panel c) denotes the synthesized beam
in the C$^{18}$O images (0$\farcs$82$\times$0$\farcs$49, P.A.=15.9$\degr$),
and that in panel f) the synthesized beam in the $^{13}$CO images
(0$\farcs$81$\times$0$\farcs$49, P.A.=15.8$\degr$).
Solid curves delineate the detected $S$-shaped feature.
Dashed lines in panel a) show the major (N-S) and minor (E-W) axes of the circumbinary disk,
and the off-center minor axes (ES-WS \& EN-WN), which are the cut lines of the Position-Velocity
diagrams shown in Figure \ref{c18opvs}.
\label{c18och3}}
\end{figure}

To help in visualizing the spatial distribution of the newly detected high-velocity components,
in Figure \ref{c18och3} (upper panels) we show maps of the high-velocity ($\gtrsim$2 km s$^{-1}$)
blueshifted (blue contours) and redshifted (red contours) C$^{18}$O (3-2) emission
superposed on the 0.9-mm dust-continuum image (grey scale).
These high-velocity components also are detected in $^{13}$CO (3-2), CS (7-6),
and SO (7$_8$-6$_7$) (no emission was detected in HC$^{18}$O$^{+}$).
For example, in Figure \ref{c18och3} (lower panels), we show the corresponding
channel maps in $^{13}$CO (see the entire velocity channel maps of the $^{13}$CO emission
in Appendix A).
At the highest blueshifted velocities (left panels), both the C$^{18}$O and $^{13}$CO emissions
peak between Sources A and B. On the other hand, at the highest redshifted velocities (left panels),
both the C$^{18}$O and $^{13}$CO emissions peak to the south-east of Source A.
At slightly lower velocities (middle panels),
the blueshifted C$^{18}$O and $^{13}$CO emissions
peak to the north from the midpoint between Sources A and B and curls northeast, 
whereas the redshifted C$^{18}$O and $^{13}$CO emissions
peak to the southeast of Source A and curls to the southwest.
The trajectory of the emission peaks at blueshifted and redshifted velocities follows
an $S$-shaped curve as highlighted by the yellow curves.
At even lower velocities (right panels), the emission in both C$^{18}$O and $^{13}$CO
now encompasses the circumbinary disk as traced in the continuum.

Figure \ref{mom1s}a shows the intensity-weighted mean velocity map in C$^{18}$O.
Whereas the velocity field measured in our previous SMA observations
at a lower angular resolution and sensitivity
can be satisfactorily modeled as pure Keplerian rotation (Figure \ref{mom1s}b), the velocity field
measured in our ALMA observation shows clear deviations from Keplerian rotation.
If the rotation was purely circular,
the line-of-sight (LOS) velocity would be symmetric about the major axis of the disk
(tilted vertical dashed lines in Figure \ref{mom1s}) and hence the ridge of peak LOS velocities
aligned with the major axis.
Even in the case of aligned Keplerian elliptical orbits, the ridge of peak LOS velocities
should lie along the straight line for any projection in the sky.
The ALMA mean-velocity map shows, however,
that this ridge deviates from the major axis in a systematic way that cannot be explained as either
circular or aligned elliptical orbits in a Keplerian disk.
In addition, a velocity gradient is apparent along the minor axis
of the disk, such that the eastern side of the disk is redshifted and the western side blueshifted.

In Figure \ref{mom1s}c, we show the residual mean-velocity map
after subtracting the best-fit Keplerian model derived from our SMA observations ($i.e.$, Figure \ref{mom1s}b)
from the mean velocity map observed with ALMA (Figure \ref{mom1s}a).
Details of the subtraction process are described in Appendix B.
For comparison, we overlay the ALMA 0.9-mm dust continuum image
after subtracting the circumstellar disks so as to highlight features in the circumbinary disk (white contours).
The subtracted mean-velocity map shows that, away from the major axis of the circumbinary disk,
the LOS velocities are, in general (except relatively close to the protostellar system),
more blueshifted to the north and more redshifted to the south than expected if the rotation
was purely Keplerian (and circular).
Along the purple and red curves drawn in Figure \ref{mom1s}c,
coinciding with a portion of the individual $U$-shaped features seen in the continuum,
the LOS velocities are $\gtrsim$0.5 km s$^{-1}$ more blueshifted and redshifted, respectively,
than are expected for pure (circular) Keplerian rotation.
Just beyond the protostellar components, the subtracted
mean-velocity map shows a patch of residual redshifted velocities to the north and blueshifted
velocities to the south. These patches indicate much lower LOS velocities than Keplerian in the close
vicinity of the binary system, implying significant changes in orbital velocities close to and
within the Roche lobes ($\lesssim$1.0$\arcsec$; see section 4.1.).

Figure \ref{c18opvs} ($Left$) shows the observed
Position-Velocity (P-V) diagrams of the C$^{18}$O emission
along the major and minor axes of the circumbinary disk
(N-S and E-W in Figure \ref{c18och3}a), as well as along the
northern and southern sides of the circumbinary disk
parallel to the minor axis (EN-WN and ES-WS in Figure \ref{c18och3}a).
Along the major axis of the circumbinary disk (N-S), the observed
P-V diagram show primarily Keplerian rotation (green curves
from the Keplerian model derived in our SMA observations).
Along the minor axis passing through the disk center (E-W),
however, the P-V diagram is not symmetric as would be expected for Keplerian rotation.
Instead, the P-V diagram along this axis shows blueshifted emission extending to the
highest velocities ($<$-2 km s$^{-1}$ from the systemic velocity)
in the west and redshifted emission extending to the highest velocities
($>$+2 km s$^{-1}$ from the systemic velocity) in the east.
Although the P-V diagram therefore indicates a blueshift to redshift velocity gradient
from west to east along the minor axis passing through the disk center,
the P-V diagram parallel to the minor axis but to the south of the disk center (ES-WS)
shows that the emission increases
in redshifted velocity to the west (red dashed line) rather than to the east.
In addition, the highest redshifted velocity is located to the west (arrow) rather than the east.
In the case of circular Keplerian rotation, the P-V diagrams along the on- and off-center minor axes
are symmetric and do not show any velocity gradients \cite{tak13}.
In the case of a rotating disk exhibiting a radial velocity component due to infall,
the P-V diagram exhibits a velocity gradient along the minor axis reflecting the
infalling motion \cite{yen10,tak13}, but the sense of the velocity gradient is the same for all
P-V diagrams parallel to the minor axis.
Thus, the different signs of the velocity gradients along and parallel to the minor axis
cannot be simply explained by combination of rotation and infall.

\placefigure{c18och}
\placefigure{mom1s}
\placefigure{c18opvs}

\section{Discussion}
\subsection{Nature of the Circumbinary Disk around L1551 NE}

Thanks to the higher angular resolution and sensitivity afforded by ALMA,
we have been able to detect internal structures in the circumbinary dust disk and
deviations from Keplerian motion in the circumbinary gas disk of L1551 NE
not previously seen in our observations with the SMA. In the following, we discuss
the physical causes of these newly-detected features. As will become apparent,
the explanation for these features provides insights into the mechanisms that
drive the exchange of angular momentum in the circumbinary disk of L1551 NE
so as to permit infall through this disk.

In the Roche potential of binary systems, circumbinary orbits close to the
binary system experience strong gravitational torques,
and cannot be described by circular or elliptical orbits
as those around single stars.
Instead, torques that impart angular momentum to the circumbinary disk create shocks
where matter that gained angular momentum collides supersonically with
matter located downstream having less angular momentum.
Theoretical simulations show that these shocks form a two-arm spiral pattern
that co-rotates with the binary system.
One arm extends from the primary through the L3 Lagrangian point in the
Roche potential, and the other arm from the secondary through the L2 point
(see Figure \ref{modelvel}a). These Lagrangian points correspond to locations
where circumbinary material flows into the Roche lobes of the respective binary components \cite{art96};
note, however, that infall does not occur along the spiral arms, but rather in regions between the spiral arms
as we will explain later.
Although the degree of density enhancement and opening angle of the spiral arms
depend on the binary parameters and physical conditions (density and temperature) in the circumbinary disk,
the two-arm spiral density pattern
is a generic prediction of all published theoretical models of
protobinary systems surrounded by a circumbinary disk \cite{art96,ba97b,bat00,gun02,och05,han10}.

To model the internal structure and dynamics of the circumbinary disk
in the specific case of L1551 NE, we performed a hydrodynamic simulation using
adaptive mesh refinement (AMR) code SFUMATO \cite{mat07}.
The relevant parameters adopted in our simulation are summarized in Table 2.
We assumed that the circumbinary disk (indeed, all the material beyond the two protostars)
is isothermal at a temperature of 13.5 K, and that its equatorial plane is aligned with the orbital
plane of the binary system. Material from the circumbinary disk therefore accretes onto the individual
circumstellar disks of the binary protostars along the same plane.
We assumed that the binary system has a circular orbit,
and adopted a total mass for the binary of 0.8 $M_{\odot}$, a mass ratio of 0.19,
and a binary orbital separation 145 AU as determined from our previous SMA observations (Table 2).
The sizes of the Roche lobes of the primary and secondary are thus $\sim$150 AU and $\sim$70 AU,
respectively (see Figure \ref{modelvel}a).
Gas is injected at the cylindrical computational boundary located at 1740 AU from the
center of the mass of the binary system. The specific angular momentum of the injected gas
($\equiv j_\mathrm{inj}$) is chosen to have a centrifugal radius of 300 AU
($i.e.$, $j_\mathrm{inj}$=461.5 AU km s$^{-1}$).
This radius corresponds to the outer radius of the circumbinary disk,
where the gas motion changes from infall in the surrounding envelope to
the Keplerian rotation in the circumbinary disk as
found in our previous SMA observations \cite{tak13}.
The choice of specific angular momentum is guided also by the need to
reproduce the inner depression in dust emission in the circumbinary disk.
As previously shown by Bate \& Bonnell (1997), the size of the inner depression in the
circumbinary disk is dependent on the specific angular momentum of the injected gas.
Our previous SMA observations show that the mass-infall rate from the surrounding
envelope to the circumbinary disk is of the order of $\sim$10$^{-6}$ $M_{\odot}~yr^{-1}$ \cite{tak13}.
In our model, we assume a density for the injected gas of 1.5 $\times$ 10$^{5}$ cm$^{-3}$ (Table 2), 
which give a comparable mass-infall rate from the envelope to the circumbinary disk of
3.3 $\times$ 10$^{-6}$ $M_{\odot}~yr^{-1}$.
We adopted the simulation result after the $\sim$20 orbital periods of the binary
($\sim$4 $\times$ 10$^{4}$ $yr$). By that stage of the simulation,
the gas distribution and kinematics have settled into a quasi-steady state
and the circumbinary disk developed spiral arms, but the accreted masses
onto the protostellar binaries have not yet changed their individual masses significantly.

Having computed the gas and also dust (by simply assuming a gas-to-dust mass ratio of 100)
distribution from our hydrodynamic simulation, we then performed radiative transfer calculations
to produce the theoretically-predicted 0.9-mm dust-continuum and C$^{18}$O images.
To better capture the intensity distribution of both the gas and dust,
when computing the radiation transfer
we assumed a combination of two power-law temperature distributions
centered, respectively, on Sources A and B.
The parameters of the individual temperature profiles are listed in Table 2.
We first fixed the temperature at a radius of 300 AU, the outer radius of the
circumbinary disk, to approximately reproduce the observed
0.9-mm continuum and C$^{18}$O line intensity peaks
in the circumbinary disk. Then, the power-law index of the temperature
profile was determined to
closely reproduce the observed flux density of the compact
continuum component (circumstellar disk) associated with each protostar.
The choice of the power-law index is controlled by the
circumstellar-disk components, and we confirmed that changing the power-law index
within the reasonable value (0 -- 0.5) does not affect the appearance of the spiral-arm
features in the circumbinary disk much, and thus the main points of the present paper.
We also note that
the use of power-law temperature profiles is in conflict with our model computations
which assume that the matter around both protostars is isothermal.
However, the adopted power-law index results in a change in temperature
of only a factor of 1.4 through the circumbinary disk
(from 33 K at $r \sim$50 AU to 23 K at $r \sim$300 AU).
The peak brightness temperatures of the $^{13}$CO and C$^{18}$O image cubes
on the northern blueshifted side of the circumbinary disk are 28.9 K and 19.8 K,
and those in the southern redshifted side 25.9 K and 24.7 K, respectively.
On the assumption of local thermodynamic equilibrium (LTE) and $X$ ($^{13}$CO) / $X$ (C$^{18}$O) = 7.7 \cite{wr94},
the excitation temperatures and optical depths can be estimated from these isotopic lines.
We find, for C$^{18}$O, an excitation temperature of $\sim$36 K and optical depth
of $\sim$1.2 at the peak brightness on the northern side, and an excitation temperature
of $\sim$33 K and optical depth of $\sim$3.0 at the peak brightness temperature on the southern side.
Thus, the C$^{18}$O emission is close to or completely optically thick at the emission peaks,
and the derived excitation temperatures at these locations correspond to
the gas temperatures at the surface of the circumbinary disk.
More typically, the peak brightness temperature of C$^{18}$O in the channel maps range
from $\sim$15 K to $\sim$20 K, suggesting that the C$^{18}$O emission is mostly optically thin.
The adopted temperature of 13.5 K in our hydrodynamic simulations is lower than
the typical peak brightness temperature measured in C$^{18}$O,
but may be more representative of the bulk of both the gas and dust in the circumbinary disk
that is concentrated close to the midplane as discussed in section 3.1.
Thus, the different temperatures used in our hydrodynamic simulation and in the radiative transfer
calculations may not present as severe an internal inconsistency as might appear on first sight.
As a check, we confirmed that using a temperature of 25 K does not produce any significant
differences in the gas distributions in our hydrodynamic simulations.
Following the radiative transfer calculations, we used the CASA task ``simobserve'' to
create the simulated visibility data for the model images with the same antenna configuration,
hour angle coverage, bandwidth and frequency resolution,
integration time, and noise level as those of the real ALMA observation.
We further performed flagging of the simulated data to match the simulated data with the real processed data,
and then made simulated theoretical images with the same imaging methods as described in Section 2.
More details of our theoretical simulations are described in Appendix B.

Figure \ref{contall}a shows the model continuum image
($i.e.$, before passing through the ALMA simulator).
This image clearly shows two spiral arms,
along with a depression or gap between the spiral arms and the two circumstellar disks.
The northern arm labeled A connects to the circumstellar disk around Source A,
and that labeled B to the circumstellar disk around Source B.
In Figures \ref{contall}b and \ref{contall}c, the simulated theoretical continuum image ($i.e.$, after the model image
has been passed through the ALMA simulator) and the observed ALMA 0.9-mm continuum
image (same as that in Figure \ref{cont}c) are shown, respectively.
The observed $U$-shaped features to the north and south of the binary system
coincide nicely with the spiral arms. Furthermore, the simulated
theoretical continuum image closely reproduces the inner depression observed in the ALMA continuum image.

Figure \ref{modelch} shows the simulated theoretical velocity channel maps
of the C$^{18}$O (3-2) line (contours)
superposed on the simulated theoretical 0.9-mm dust-continuum image (grey scale).
These channel maps closely capture all the important features seen in
the ALMA image cube as described above. At the highest blueshifted velocities of
2.8 -- 4.3 km s$^{-1}$, the C$^{18}$O emission is located between Sources A and B.
At lower blueshifted velocities of 4.6 to 5.2 km s$^{-1}$, the C$^{18}$O emission
extends towards the north before curling northeast from the midpoint between Sources A and B.
By comparison, at the highest redshifted velocities of 9.9 -- 10.3 km s$^{-1}$,
the C$^{18}$O emission is located to the southeast of Source A.
At slightly lower redshifted velocities of 8.6--9.7 km s$^{-1}$,
the C$^{18}$O emission is located to the southeast of source A and curls to the southwest.
At lower blueshifted and redshifted velocities closer to the systemic, the channel maps
primarily exhibit Keplerian rotation. The P-V diagrams
constructed from the simulated theoretical images are
shown in Figure \ref{c18opvs} ($Right$), and closely reproduce
the velocity features in the observed P-V diagrams as described above;
$i.e.$, the velocity gradient along the minor axis, and
the different signs of the velocity gradient along the minor axis
compared to the velocity gradients along cuts parallel to the minor axis on either sides of the circumbinary disk.

The simulated theoretical C$^{18}$O mean-velocity map shown in Figure \ref{mom1s}d
shows a systematic deviation in the peak LOS velocity away from the major axis of the circumbinary disk.
The ridge in peak LOS velocities traces an $S$-shaped pattern as highlighted by the yellow line in
Figure \ref{mom1s}d that is seen also
in the velocity field of our ALMA C$^{18}$O map (Figure \ref{mom1s}a).
This deviation results in higher blueshifted LOS velocities at the north-eastern
(purple curve in Figure \ref{mom1s}c) and higher redshifted LOS velocities at the
south-western (red curve) portions of the circumbinary disk than predicted for Keplerian rotation.
These portions of the circumbinary disk correspond, respectively,
to the eastern part of Arm A and western part of Arm B.
Because the spiral arms are produced by gravitational torques
from the binary protostars that impart angular momentum onto matter in the circumbinary disk,
the matter along the arms orbits faster than Keplerian and expands outward,
thus producing a peak in the LOS velocity (see Figure \ref{modelvel}).
The east (blueshifted) to west (redshifted)
velocity gradients along the offset minor axes seen in the observed and theoretical P-V diagrams
(Figure \ref{c18opvs}) reflect the expanding motion of the arms,
since the disk near-side is on the eastern side and the far-side on the western side.
On the other hand,
the minor axis of the circumbinary disk axis does not cross the spiral arms, 
but instead crosses regions where our theoretical model predicts torques largely extract angular
momentum from the circumbinary disk (Figure \ref{modelvel}).
Here, our theoretical model predicts infall occurs, thus explaining the observed velocity gradient
along the minor axis of the circumbinary disk. As the observed velocities on both sides of center
along the minor axis reach $\pm$0.5 km s$^{-1}$, the infall velocity corresponds to
$\sim$0.5 km s$^{-1}$ / $\sin i$ $\sim$0.6 km s$^{-1}$ (where $i$ is the inclination of the disk).
The inferred infall velocity closely matches that predicted in our model.

\begin{figure}
\epsscale{0.8}
\plotone{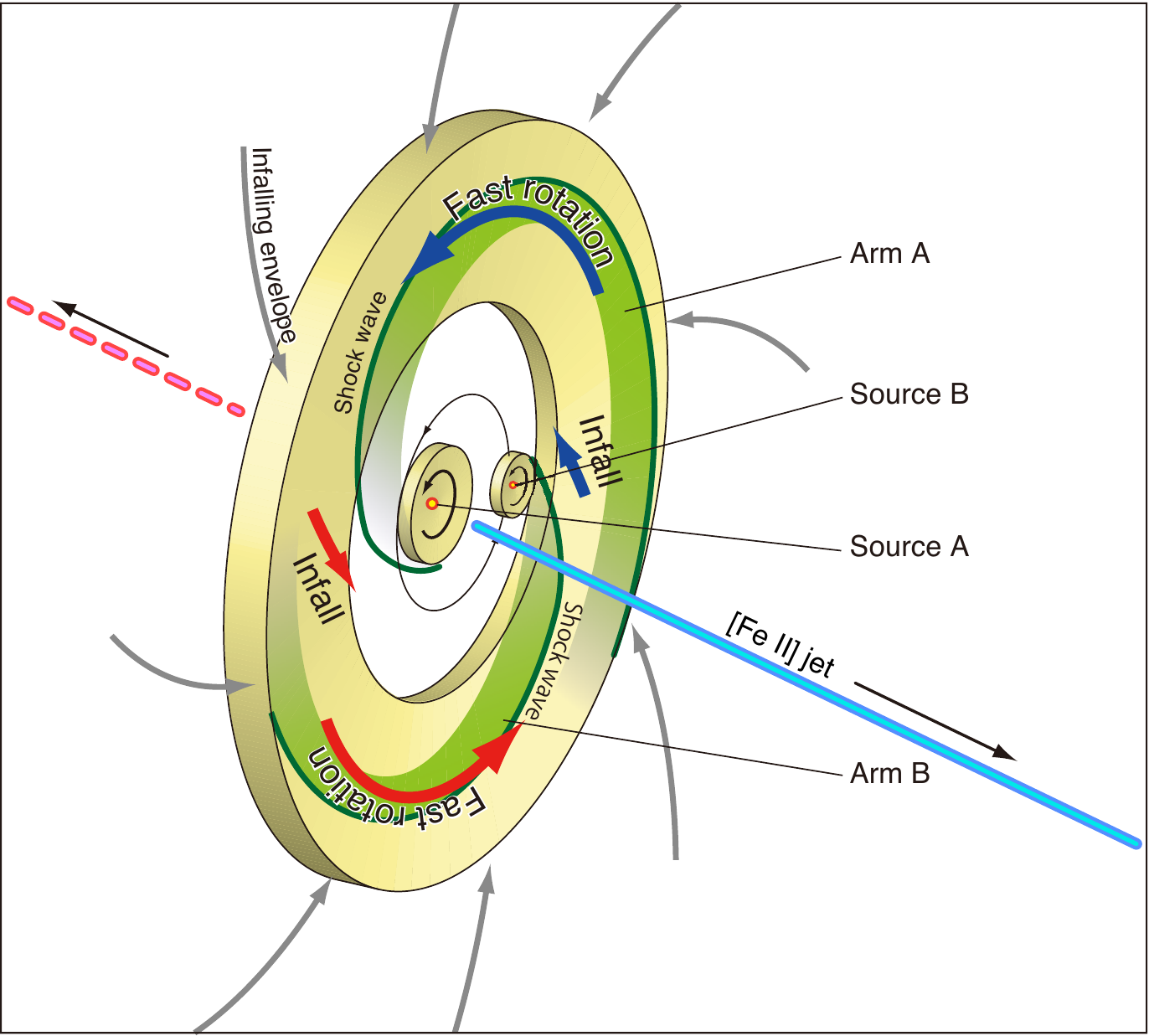}
\caption{Schematic picture of the protostellar binary system L1551 NE
revealed with the present ALMA observation.
\label{scheme}}
\end{figure}

In summary, both the $U$-shaped dust brightenings in the circumbinary disk of L1551 NE
and its velocity field as measured in C$^{18}$O can be explained solely by the effects of
gravitational torques from the binary components.
The gravitational torques, rather than
magnetic fields or other mechanisms for generating turbulence, constitute
the primary driver for exchanging angular momentum in the circumbinary disk
so as to permit infall through this disk and hence growth of L1551 NE.
In Figure \ref{scheme}, we present a schematic picture
that shows the major features so-far discovered in our study of L1551 NE.

\subsection{Binary Accretion}

An important and as yet unresolved issue in binary star formation is, once one protostar
gains a significantly larger mass than the other, which protostar accretes at a higher rate
and what factors determine the mass ratio of the final binary product.
At approximately solar masses, observations suggest that binary stars do not exhibit
a preferred mass ratio.
Duquennoy \& Mayor (1991) showed that the mean mass ratio of the solar-type stars is $\sim$0.4,
and Reid \& Gizis (1997) found that the mass ratio distribution is approximately flat from $\sim$0.1 to 1.
Theoretical simulations have produced conflicting results on whether the primary or secondary
protostar accretes at a higher rate, and thus whether the mass ratio of binary systems is driven
away from or towards unity.
Bate (1997), Bate \& Bonnell (1997), and Bate (2000) suggest that
the secondary protostar accretes at a higher rate because it is located further away
from the center of the mass of the binary system, and thus sweeps up more material in its orbit
than the primary protostar.
On the other hand, numerical simulations by Ochi et al. (2005)
and Hanawa et al. (2010) employing higher resolutions
show that the accretion rate of the primary is larger than that of the secondary.
They argue that high-resolution simulations are necessary to properly follow the complex
gas motions around the L2 and L3 Lagrangian points where material from the circumbinary disk
flows into the Roche lobes. Similar results are also obtained by Fateeva et al. (2011).

We anticipate that observations that can properly resolve the gas kinematics
around the L2 and L3 Lagrangian points and within the Roche lobes
will be crucial for addressing the factors that determine the relative
accretion rates onto the individual components of a binary system.
Future observations with ALMA will be able to resolve the gas kinematics
around the Lagrangian points and within the Roche lobes, and to provide
important observational insights on the mass accretion onto the primary
and secondary of the protostellar binaries.


\section{Summary}

We have observed the Class I binary protostellar system L1551 NE
in the 0.9-mm dust continuum and C$^{18}$O (3-2) line with ALMA during Cycle 0.
Compared with our previous observations of L1551 NE in the same tracers with the SMA,
the improvement in angular resolution is a factor of $\sim$1.6 (in beam area)
and the sensitivity a factor of $\sim$6 (in brightness temperature).
We find the following new features in our ALMA observation that were either not detected
or revealed in our previous SMA observations:

1. The central weakly-extended continuum component detected with the SMA is split into two compact
(unresolved) components corresponding to the circumstellar dust disks of the individual protostars.
Assuming a gas-to-dust mass ratio of 100, $\kappa_{0.9mm}$ = 0.053 cm$^{2}$ g$^{-1}$,
and $T_{d}$ = 10 -- 42 K,
we derive a mass of $\sim$0.005 - 0.044 $M_{\odot}$
for the circumstellar disk of Source A (primary) and $\sim$0.003 - 0.022 $M_{\odot}$ for Source B (secondary).

2. The dust image also reveals $U$-shaped brightening on the southern side of the circumbinary disk,
and emission depression or gaps between the circumbinary disk and the circumstellar disk of the primary protostar
(Source A, located to the south-east of the secondary, Source B).
After subtracting the two circumstellar disks, we find that protrusions to the north of the two protostars
and brightenings on the northern side of the circumbinary disk comprise yet another $U$-shaped brightening.
Assuming as before a gas-to-dust mass ratio of 100, $\kappa_{0.9mm}$ = 0.053 cm$^{2}$ g$^{-1}$,
and $T_{d}$ = 10 -- 42 K,
we derive a mass of $\sim$0.007 - 0.059 $M_{\odot}$ for the circumbinary disk.

3. The C$^{18}$O emission is detected to
higher blueshifted velocities of 2.8 -- 4.3 km s$^{-1}$ and redshifted velocities
of 9.0 -- 10.3 km s$^{-1}$ than in our previous observations with the SMA.
Whereas emission at lower blueshifted and redshifted velocities in the circumbinary disk
(as previously seen with the SMA) primarily trace Keplerian rotation,
the higher velocity emission exhibits systematic deviations from Keplerian motion.
The mean velocity map of the circumbinary disk exhibits an $S$-shaped pattern in its peak line-of-sight velocity.
Furthermore, there is an east-west (redshifted to blueshifted) velocity gradient along the minor axis
of the circumbinary disk. By contrast, parallel to the minor axis to  the north and south of the center,
the circumbinary disk exhibits a velocity gradient having an opposite sense to that along the minor axis.
Unlike in the continuum, there is no appreciable depression in C$^{18}$O between the
circumbinary disk and circumstellar disk of the primary protostar, suggesting that this region is
not devoid of matter (implying also that C$^{18}$O is more optically thick than the dust continuum at 0.9 mm).


To interpret the newly-observed features unveiled with ALMA, we performed a hydrodynamic simulation of
a binary system embedded in a circumbinary disk that is tailored to the inferred properties
of L1551 NE. The model does not include magnetic fields or any artificial injection of turbulence
(to help transfer angular momentum), but simply considers the effects of gravitational torques
from the central binary system on the surrounding matter.
To generate simulated theoretical images for direct comparison with the observed images,
we computed the transfer of radiation through the circumstellar disk associated with
each protostar as well as the circumbinary disk, and passed the resultant model images
through the ALMA observing simulator.
A comparison between the observed and theoretical images reveals that:

4. The observed $U$-shaped features on opposing sides of the circumbinary disk
correspond to the predicted pair of spiral arms generated by gravitational torques
from the central binary system. These spiral arms constitute density enhancements
created by shocks where gravitational torques deposit angular momentum to the
circumbinary disk. The northern arm is connected to Source A and the southern arm
to Source B, with both arms co-rotating with the binary system.

5. The observed velocity field of the circumbinary disk
($e.g.$, $S$-shaped pattern in the mean-velocity map)
can be explained entirely by the motion of matter in the Roche potential of a binary system.
The observed velocity gradients parallel to the minor axis of the circumbinary disk but
displaced both north and south of the disk center so as to pass through the spiral arms
reflect expanding gas motion where gravitational torques impart angular momentum
to the circumbinary disk.
On the other hand, the observed opposite sense in velocity gradient along the minor axis
of the circumbinary disk that does not pass through the spiral arms reflects infalling gas motion
where gravitational torques extract angular momentum.
The infall velocity of $\sim$0.6 km s$^{-1}$ inferred from our observation is
closely reproduced in our model, which together with all the other abovementioned
close agreement between our observation and model, suggest that gravitational torques
constitute the primary driver for exchanging angular momentum in the circumbinary disk
so as to permit infall through this disk.

Our results demonstrate that gravitational torques from the binary stars, without
additional mechanisms such as magnetic braking, can efficiently
exchange angular momentum in the circumbinary disk so as to permit infall in the circumbinary disk.
Future higher-resolution observations of the circumbinary disk in L1551 NE that are able to
resolve gas motions within the Roche lobes
promise to provide valuable information on accretion from the circumbinary disk to the individual circumstellar disks.

\acknowledgments
We are grateful to the anonymous referee for insightful comments
and suggestions. We would like to thank N. Ohashi, M. Hayashi, and M. Momose for their
fruitful discussions, and all the ALMA staff supporting this work. S.T. acknowledges a grant from
the Ministry of Science and Technology (MOST) of Taiwan
(MOST 102-2119-M-001-012-MY3)
in support of this work. J.L. is supported by the GRF
grants of the Government of the Hong Kong SAR under HKU
703512P for conducting this research.
T.M. is supported by the 
Grants-in-Aid for Scientific Research (A) 24244017 and (C) 23540270 from the Ministry of Education, Culture, Sports, Science and Technology, Japan.
This paper makes use of the following ALMA data:
ADS/JAO.ALMA\#2011.0.00619.S. ALMA is a partnership of
ESO (representing its member states), NSF (USA) and NINS (Japan),
together with NRC (Canada) and NSC and ASIAA (Taiwan), in cooperation
with the Republic of Chile. The Joint ALMA Observatory is operated by
ESO, AUI/NRAO and NAOJ.
Numerical computations were in part carried out on Cray XC30 at Center for Computational Astrophysics, National Astronomical Observatory of Japan.

\appendix
\section{Velocity Channel Maps of the $^{13}$CO ($J$=3-2) Emission}

\begin{figure}
\epsscale{0.55}
\plotone{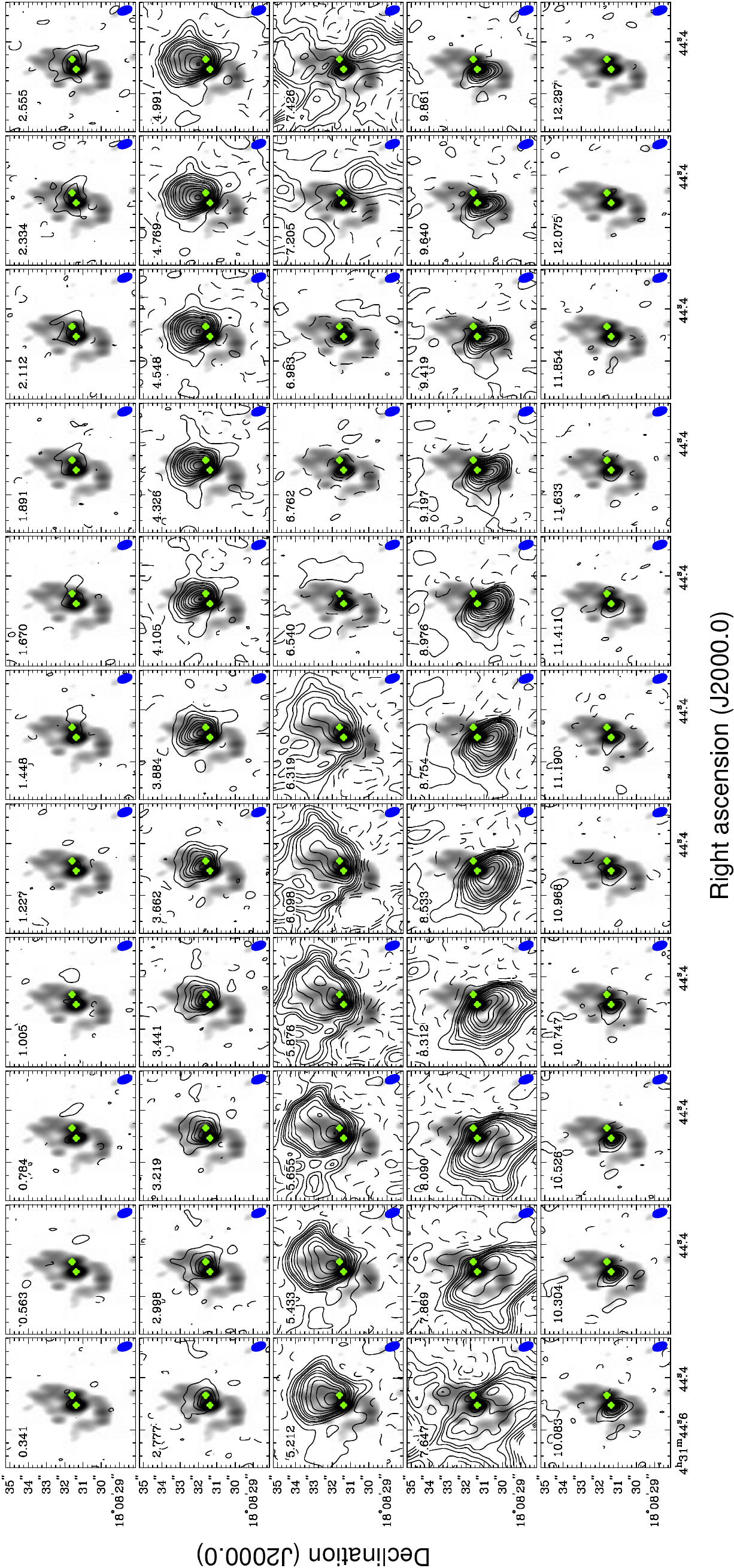}
\caption{Observed velocity channel maps of the $^{13}$CO ($J$=3-2) emission
(contours) superposed on the 0.9-mm dust-continuum emission (grey) in L1551 NE,
taken with ALMA. Numbers at the top-left corners denote the LSR velocities.
Contour levels are 3$\sigma$, 9$\sigma$, 15$\sigma$, 20$\sigma$, 30$\sigma$, 40$\sigma$,
50$\sigma$, and then in steps of 20$\sigma$ (1$\sigma$ = 7.8 mJy beam$^{-1}$).
Crosses show the positions of the protostellar binary, and
filled ellipses at the bottom-right corners the synthesized beam
(0$\farcs$86$\times$0$\farcs$51; P.A.=21.9$\degr$).
\label{13coch}}
\end{figure}

Figure \ref{13coch} shows the observed velocity channel maps of the $^{13}$CO ($J$=3-2)
emission in L1551 NE. As shown in Figure \ref{c18och3}, in the high blueshifted
and redshifted velocities the $^{13}$CO emission primarily traces the same velocity features
as those traced by the C$^{18}$O emission.
The $^{13}$CO emission peaks between Sources A and B in the highest blueshifted velocities
(0.34 -- 3.4 km s$^{-1}$), while in the highest redshifted velocities (9.9 -- 12.1 km s$^{-1}$)
to the southeast of Source A.
At blueshifted velocities of 3.7 -- 4.1 km s$^{-1}$
the C$^{18}$O emission extends northeast from the midpoint between Sources A and B,
and at redshifted velocities of 9.2 -- 9.9 km s$^{-1}$ to the south and southwest.
At velocities of 4.3 -- 5.4 km s$^{-1}$ and 7.9 -- 9.0 km s$^{-1}$,
the $^{13}$CO emission is located to the northwest and southeast of the protobinary,
respectively, which traces the Keplerian rotation of the circumbinary disk.
Around the systemic velocity (5.7 -- 7.6 km s$^{-1}$)
the $^{13}$CO emission is complicated and affected by the effect of the missing flux.

\section{Subtraction of the Keplerian Rotation Motion}

In our previous study of L1551 NE with the SMA, we found that the bulk motion
of the circumbinary disk is Keplerian (circular).
To separate systematic deviations from the bulk Keplerian motion as seen in our ALMA observation,
we subtracted the global Keplerian rotation from the observed C$^{18}$O (3-2) velocity channel maps
in the manner described below.

Assuming that the circumbinary disk is geometrically thin
the observed line-of-sight velocities ($\equiv v_{obs}$) of the molecular gas on the disk plane can be
expressed as
\begin{equation}
v_{obs} (\alpha,\delta)
= v_{sys} + v_{\phi} (\alpha,\delta) \sin(i) \cos(\Phi-\theta) + v_{rad} (\alpha,\delta) \sin(i) \sin(\Phi-\theta).
\end{equation}
In the above expressions, ($\alpha$, $\delta$) are the coordinates
in right ascension and declination with respect to the disk center,
$\Phi$ the azimuthal angle from the major axis on the disk plane,
$v_{sys}$ the systemic velocity,
$v_{\phi}$ and $v_{rad}$ the azimuthal and radial velocities, respectively,
$i$ the inclination angle (= 62$\degr$), and $\theta$ the position angle (= 167$\degr$).
The azimuthal velocity ($v_{\phi}$) can be expressed as a combination of a Keplerian
($v_{kep}$) and non-Keplerian ($v_{\phi}^{nonkep}$) components;
\begin{equation}
v_{\phi} (\alpha,\delta)=v_{\phi}^{nonkep}(\alpha,\delta)+v_{kep}(r).
\end{equation}
The Keplerian velocity is expressed as,
\begin{equation}
v_{kep}(r)=\sqrt{\frac{GM_{\star}}{r}},
\end{equation}
where
\begin{equation}
r = \sqrt{(\frac{x}{\cos i})^2 + y^2},
\end{equation}
\begin{equation}
x = \alpha \cos(\theta) - \delta \sin(\theta),
\end{equation}
\begin{equation}
y = \alpha \sin(\theta) + \delta \cos(\theta),
\end{equation}
$r$ is the radius, $x$ and $y$ are the coordinates in the disk plane along the minor and major axes,
$G$ is the gravitational constant, and $M_{\star}$ is the total mass of the binary (= 0.8 $M_{\odot}$).
Thus, at each position ($\alpha$, $\delta$), the line-of-sight velocity after subtraction of the
bulk Keplerian motion can be expressed as
\begin{equation}
v_{obs}^{non-kep} (\alpha,\delta) = v_{obs}(\alpha,\delta) - v_{kep}(r)\sin(i) \cos(\Phi-\theta) - v_{sys}.
\end{equation}

We computed the line-of-sight non-Keplerian velocity component for each pixel
in the channel maps using expression A7,
and then resampled the image cube to create new velocity channel maps,
as well as the new mean velocity map (Figure \ref{mom1s}c).

\section{Theoretical Model of L1551 NE}

\subsection{Model of the Hydrodynamic Simulation}

To help interpret the two $U$-shaped brightenings and velocity field of the
circumbinary disk in L1551 NE as observed with ALMA, we
performed a hydrodynamic simulation of an accreting binary system tailored
to the specific properties of L1551 NE.
In our model, we assume that the binary stars have a circular orbit.
The masses and separation of the protostars are as inferred from our
previous observations of L1551 NE with the SMA, corresponding to
a mass of 0.675 $M_{\odot}$ for Source A, mass of 0.125 $M_{\odot}$
for Source B, and a separation of $D$=145 AU.
Using cylindrical coordinates $(R, \varphi, z)$,
we placed the protostars at the midplane (z=0 plane) such that their center
of mass coincides with the origin (R=0).
The computational domain is defined as 
$0 \leq R \leq R_\mathrm{bound}$, $0\leq\varphi\leq 2\pi$, and 
$-Z_\mathrm{bound}/2 \leq z \leq Z_\mathrm{bound}/2$, where
$R_\mathrm{bound} = Z_\mathrm{bound} = 12D = 1740$~AU.
Gas having a constant density is injected at the cylindrical boundary
at $R=R_\mathrm{bound}$ with a velocity
\begin{equation}
\left(
\begin{array}{c}
v_R \\
v_\varphi \\
v_z
\end{array}
\right)
= 
\left(
\begin{array}{c}
\displaystyle
- \sqrt{\frac{2GM_\star}{R_\mathrm{bound}} - \left(\frac{j_\mathrm{inj}}{R_\mathrm{bound}} \right)^2 }\\
\displaystyle
\frac{j_\mathrm{inj}}{R_\mathrm{bound}}\\
0
\end{array}
\right).
\end{equation}
The radial velocity corresponds to the infall velocity for
material experiencing free-fall from infinity in the mid-plane.
The specific angular momentum of the injected gas is given by
\begin{equation}
j_\mathrm{inj} = \sqrt{ GM_\star R_\mathrm{cent} },
\end{equation}
where
$R_\mathrm{cent}$ ($=300$~AU) is the centrifugal radius of the gas
as determined from our observations with the SMA \cite{tak12,tak13}.
The mass-injection rate into the computational domain depends on the 
density of the injected gas, and is given by
\begin{equation}
\dot{M} = 
- 2 \pi R_\mathrm{bound} Z_\mathrm{bound}
\rho_\mathrm{inj} v_R
= 2.22 \times 10^{-6} 
\left(
\frac{n_\mathrm{inj}}{10^5 \,\mathrm{cm}^{-3}}
\right)
\, M_\odot \mathrm{yr}^{-1}, 
\label{eq:accretionrate}
\end{equation}
where $\rho_\mathrm{inj}$ and $n_\mathrm{inj}$ denote
the mass density and the number density, respectively, of the injected gas.
For simplicity, we assume that the gas is isothermal throughout the simulation
with a corresponding sound speed of
$c_s = 0.1 \sqrt{GM_\star/D} = 0.221$~km~s$^{-1}$. 
The gas temperature is therefore 13.5~K, a value that is typical for
dense molecular cloud cores and infalling envelopes.
%
The self-gravity of
the gas can be safely ignored because, based on both our SMA and ALMA
observations, the mass of the
circumbinary disk is considerably smaller than the total mass of the
binary stars (see section 3.1).  
The gas is therefore attracted inwards solely by the gravity of the
protostars.

\subsection{Method of the Hydrodynamic Simulation}

The calculation was performed using SFUMATO, a three-dimensional AMR
code \citep{mat07}. The hydrodynamic scheme was modified to have a
third order of accuracy in space and second order in time.
The computational domain of $(2R_\mathrm{bound})^2 \times Z_\mathrm{bound}$ is resolved
by the base grid of ($l=0$) with $128^2\times 64$ cells.  In the AMR
hierarchy, the maximum grid level is set at $l=5$.  The cell width is
$\Delta x = 0.85$~AU on the finest grid of $l=5$, compared with
$\Delta x = 27$~AU on the base grid of $l=0$.
The effective resolution therefore corresponds to
$4096^2\times2048$. The initial hierarchical grid was fixed
during the calculation, and no further grid refinement during the time
integration is performed.
The two protostars are represented by two point masses that comprise
sink particles.  The sink particles accrete gas within a given radius referred to as
the sink radius.
The sink radius is set at 3.4~AU, which is
considerably smaller than the binary separation of $D = 145$~AU,
and we adopted the accretion method described by Krumholz et al. (2004).
The simulation is performed in a rotating frame where the sink particles
are at rest, and takes into account both centrifugal and Coriolis forces.

\subsection{Overview of the Evolution}
\label{sec:overview_of_the_evolution}
\begin{figure}
\epsscale{1.0}
\plotone{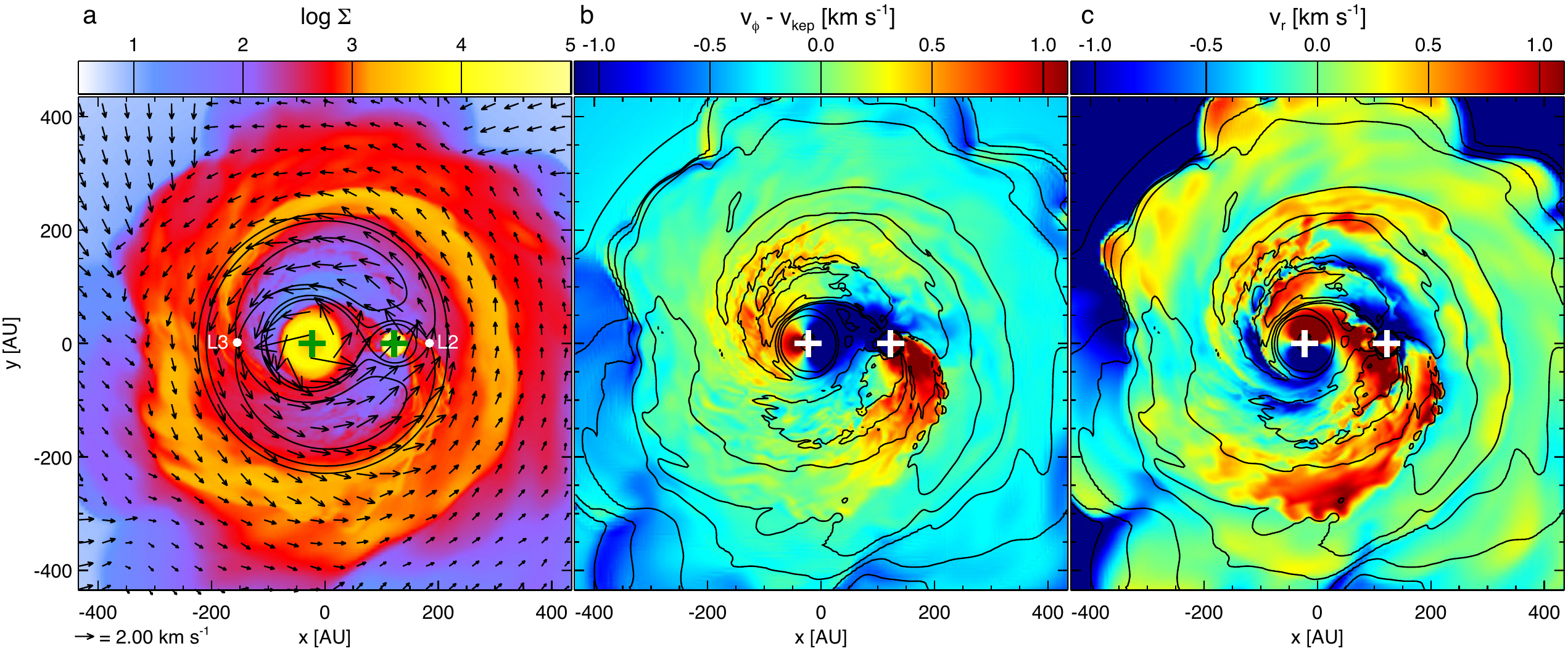}
\caption{Circumbinary disk reproduced by our AMR hydrodynamic simulation of gas
accretion onto the protostellar binary at the stage of $t = 4.13\times10^4$~yr
(21.1 orbital periods of the binary).
a) Color scale denotes the logarithmic surface density distribution.
The surface density is normalized so that the binary separation and
the gas density at the outer boundary are unity.  
Arrows denote the gas velocity in the mid-plane of the protostellar binary
system.  The Roche potential is overplayed by contours.  Crosses
indicate the positions of the protostellar binary;
the left and right crosses correspond to Source A (primary star) and
Source B (secondary star), respectively.
b) Color scale denotes a difference between the rotation velocity and
the Keplerian velocity in the mid-plane, $v_{\phi}- v_{kep}$,  
where $v_{kep} = \sqrt{GM_\star/r}$, $M_\star$ denotes the total mass of Sources A and
B, and $r$ denotes a distance from the origin which is defined
by the center of the mass of the protostellar binary. 
The red regions exhibit faster rotation than that
expected from the Kepler's law, and vice versa for the blue regions.
The logarithmic surface density is overplayed by contours, where
the contour levels are  $\log \Sigma = 1, 0.5, 1.5, \cdots 4$.
The spiral arms tend to exhibit faster rotation.
c) Color scale denotes the radial velocity $v_r$ in the mid-plane. 
The radial velocity is measured from the origin.
The logarithmic surface density is overplayed by contours.
The spiral arms tend to exhibit outflows (red), while the inter-arms tend to exhibit inflows (blue).
Note that, in panels b) and c), extremely high values are obtained
inside and near the Roche robe because the Keplerian rotation around
the origin is expected only outside the Roche lobe.
\label{modelvel}}
\end{figure}

In the simulation, the injected gas falls toward the center, and initially forms a circumbinary disk
having an inner radius of $\sim$200 AU and an outer radius of $\sim$400 AU. At this early stage, the circumbinary disk,
which is surrounded by an infalling envelope, therefore exhibits a central gap.
Matter at the inner edge of the circumbinary disk is prevented from falling into the central gap
by its angular momentum.
After $\sim$20 orbital periods, by which time the gas has settled into a quasi-steady state,
the circumbinary disk has developed spiral arms as shown in Figure \ref{modelvel}a.
The spiral arms in the circumbinary disk are excited by
the orbiting binary protostars; the rotation of their non-axisymmetric potential
generates torques that act on the gas in the circumbinary disk.
This non-axisymmetric potential is expressed as the Roche potential drawn in Figure \ref{modelvel}a.
The spiral arms co-rotate with the protostellar binary, and sweep through the circumbinary disk
and generate the shock waves,
as their angular velocity is higher than that of the circumbinary disk (see Figure \ref{modelvel}b).
The shock wave pushes gas downstream
in the circumbinary disk. In other words, the gas in the spiral arms obtains angular momentum
which is transferred from the orbital angular momentum of the protostellar binary via the
non-axisymmetric potential and shock waves.
The higher orbital velocity in the spiral arms causes the positive radial velocity there as shown in
Figure \ref{modelvel}c. Where the angular momentum is lowered by the non-axisymmetric potential,
as indicated by regions having negative radial velocities in the circumbinary disk shown
in Figure \ref{modelvel}c, infall occurs.
The positive and negative radial velocities coincide with the downstream and upstream sides, respectively,
of the shock waves.

Infall through the circumbinary disk promotes accretion onto the
circumstellar disks through the gap in the circumbinary disk. Gas
falls into this gap from the upstream side of the spiral arms, and then
onto the circumstellar disks orbiting inside the gap. In our simulation,
we find that the mass-accretion rate onto the
secondary star (Source B) is about one order of magnitude higher than
that onto the primary star (Source A). Thus, in our simulation, the
binary mass ratio increases as the protostellar binary
evolves. Similar results have also been reported in previous
works \citep[e.g.,][]{ba97b}.

The simulation shows that the accretion rate onto Source B is
roughly equal to that onto the circumbinary disk in the quasi-steady state.
In our simulation we assumed the number density of the injected gas as
$1.5 \times 10^5 \,\mathrm{cm}^{-3}$ (see Table~2 and equation~[\ref{eq:accretionrate}]).
This assumed number density provides an accretion rate from the infalling envelope
to the circumbinary disk of $3.3 \times 10^{-6} M_\odot\,\mathrm{yr}^{-1}$, similar
to that estimated from our previous SMA observations \cite{tak13}.
By using this number density,
the radiative transfer calculation reproduces
the dust emission as shown in Figure \ref{contall}a.
This indicates that the mass accretion rates onto
Sources A and B are $\sim 10^{-7} M_\odot\,\mathrm{yr}^{-1}$ and
$\sim 10^{-6} M_\odot\,\mathrm{yr}^{-1}$, respectively.

\subsection{Radiative Transfer Calculation}

\begin{figure}
\epsscale{0.8}
\plotone{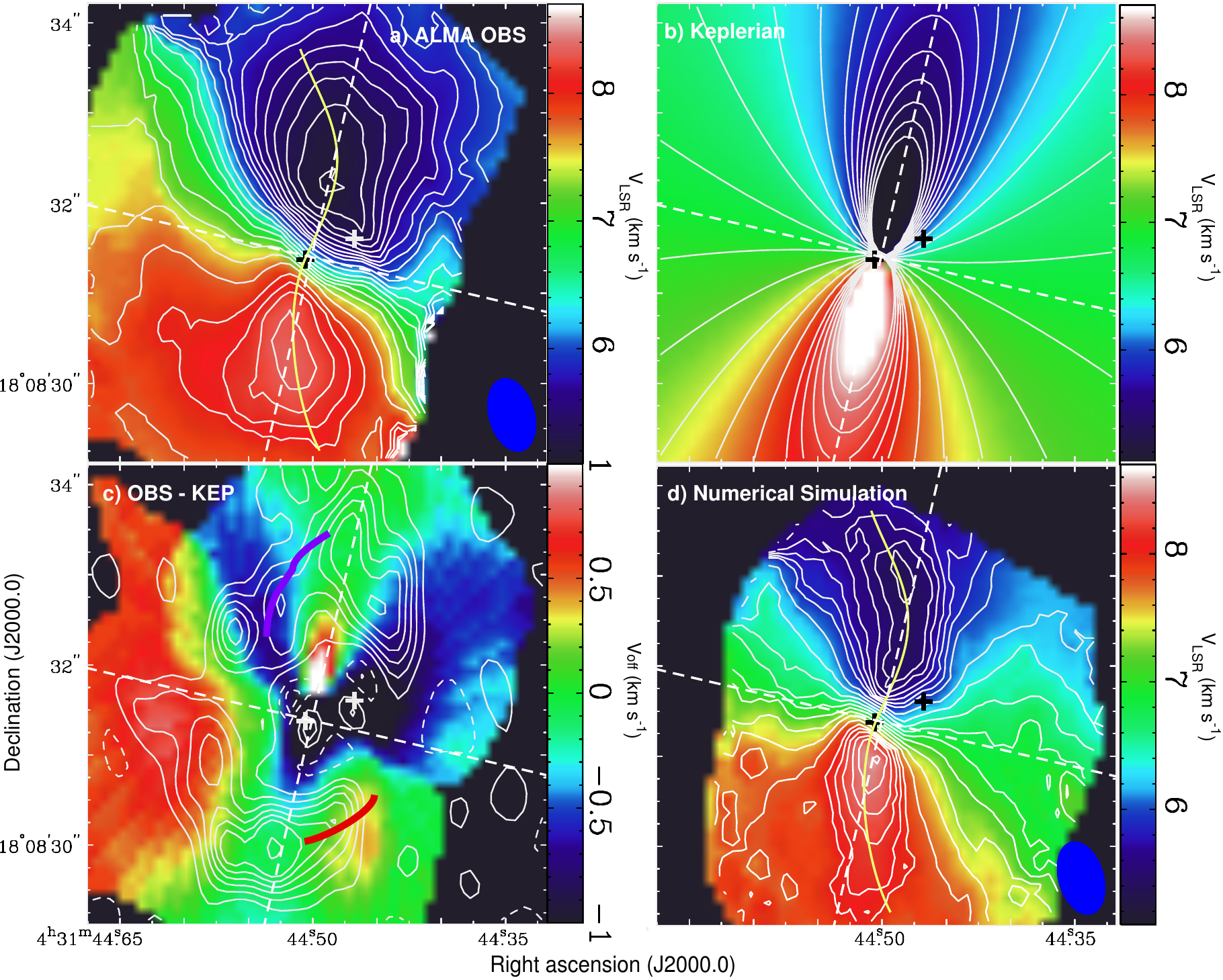}
\caption{a) Intensity-weighted mean velocity (moment 1) map of
the C$^{18}$O ($J$=3-2) emission in L1551 NE observed with ALMA.
Contour levels are from $V_{\rm LSR}$=5.2 km s$^{-1}$
(bluest contour to the north) in steps of 0.1 km s$^{-1}$ until $V_{\rm LSR}$=6.0 km s$^{-1}$,
and in steps of 0.2 km s$^{-1}$ until 8.0 km s$^{-1}$, and then in steps of 0.1 km s$^{-1}$
until $V_{\rm LSR}$=8.3 km s$^{-1}$ (reddest contour to the south).
Tilted vertical and horizontal dashed lines denote the major and minor axes of the circumbinary disk,
respectively.
Crosses show the positions of the protostellar binary, and
a filled ellipse at the bottom-right corner shows the synthesized beam
(0$\farcs$82$\times$0$\farcs$49; P.A.=15.9$\degr$).
A solid curve traces (by eye) the detected $S$-shaped
velocity pattern.
b) Moment 1 map in the case of the axisymmetric Keplerian rotation derived
from our previous SMA observations of L1551 NE.
Contour levels
are the same as those in panel a), except for the two additional red contours
of $V_{\rm LSR}$=8.4 km s$^{-1}$ and 8.5 km s$^{-1}$.
c) Map of the residual line-of-sight velocity after subtracting the
Keplerian-rotation motion (panel b) from the observed mean velocity (panel a).
White contours show the map of the observed 0.9-mm dust-continuum emission,
after subtracting the central two compact components which most likely arise
from the circumstellar disks around the binary stars (same as that in Figure \ref{cont}d).
Purple and red curves delineate the arm portions
where the observed velocities are faster than that expected from the
Keplerian rotation.
d) Theoretically-predicted moment 1 map of
the C$^{18}$O ($J$=3-2) emission in L1551 NE, calculated from our hydrodynamic
and radiative transfer calculations and the ALMA observing simulation. Contour levels
are the same as those in panel b).
\label{mom1s}}
\end{figure}

\begin{figure}
\epsscale{0.7}
\plotone{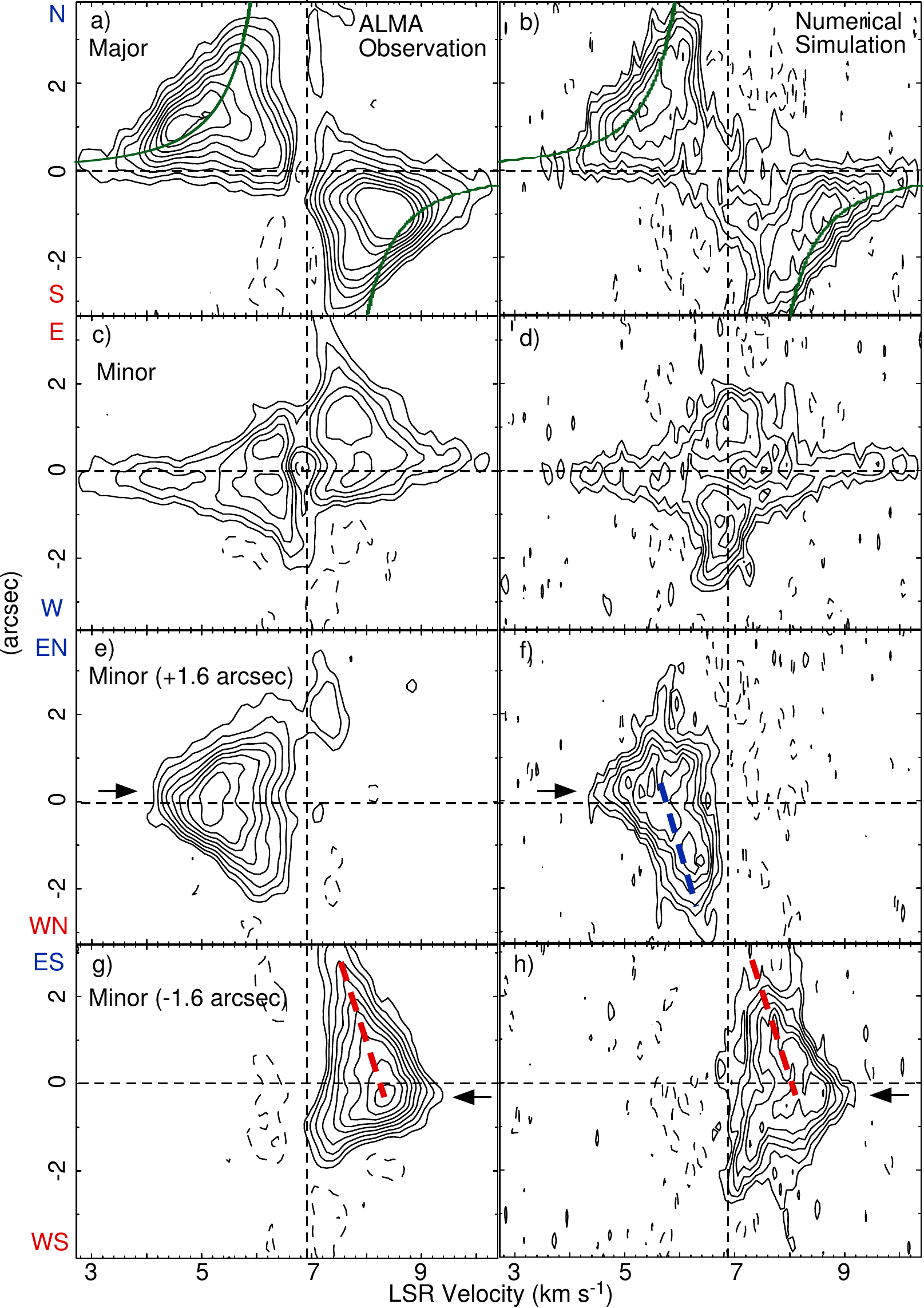}
\caption{
Observed (left panels) and model (right) Position-Velocity diagrams of the
C$^{18}$O ($J$=3-2) emission in L1551 NE along the major and minor axes
of the circumbinary disk, and along the cuts parallel to the minor axis
passing through $\pm$1.6$\arcsec$ offsets from the center along the major axis.
Contour levels are 3$\sigma$, 6$\sigma$, 10$\sigma$, and then in steps of
5$\sigma$ until 45$\sigma$ (1$\sigma$ = 15.8 mJy beam$^{-1}$).
Horizontal and vertical dashed lines denote the disk centroid positions along
the cut lines and the systemic velocity 6.9 km s$^{-1}$, respectively.
Solid green curves in panel a) and b) show the Keplerian rotation curve
derived from the previous SMA observations. Arrows in the bottom four panels
show the positional offsets of the highest velocity emission from the disk
major axis, and dashed lines the velocity gradients detected
along the off-center minor axes.
\label{c18opvs}}
\end{figure}

\begin{figure}
\epsscale{1.0}
\plotone{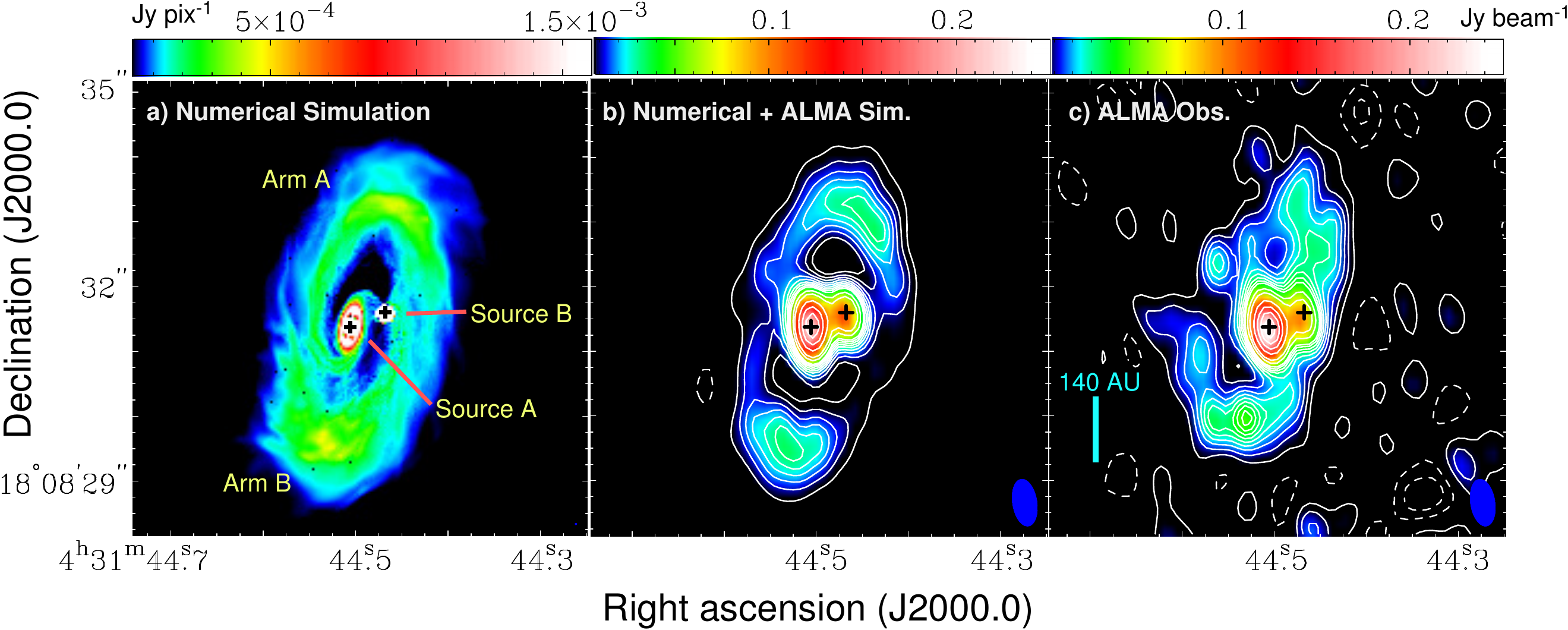}
\caption{a) 0.9-mm dust-continuum image of L1551 NE taken with ALMA. Contour levels
start from 2$\sigma$ in steps of 2$\sigma$ until 12$\sigma$, and in steps of 6$\sigma$
until 30$\sigma$, and then in steps of 15$\sigma$ (1$\sigma$ = 2.6 mJy beam$^{-1}$).
The highest contour level is 90$\sigma$.
The lower-left and upper-right crosses indicate
the positions of Sources A and B, respectively. A filled ellipse at the bottom-right corner shows the
synthesized beam (0$\farcs$72$\times$0$\farcs$36; P.A.=9.1$\degr$).
b), c) Theoretically-predicted 0.9-mm dust-continuum images of L1551 NE. We performed
the radiative transfer calculation with the gas distribution computed from our 3-D hydrodynamic
model to produce the theoretical image in panel c). Then we conducted the ALMA observing
simulation to make the theoretically-predicted ALMA image shown in panel b). Contour levels
and symbols are the same as those in panel a).
\label{contall}}
\end{figure}

\begin{figure}
\epsscale{0.8}
\plotone{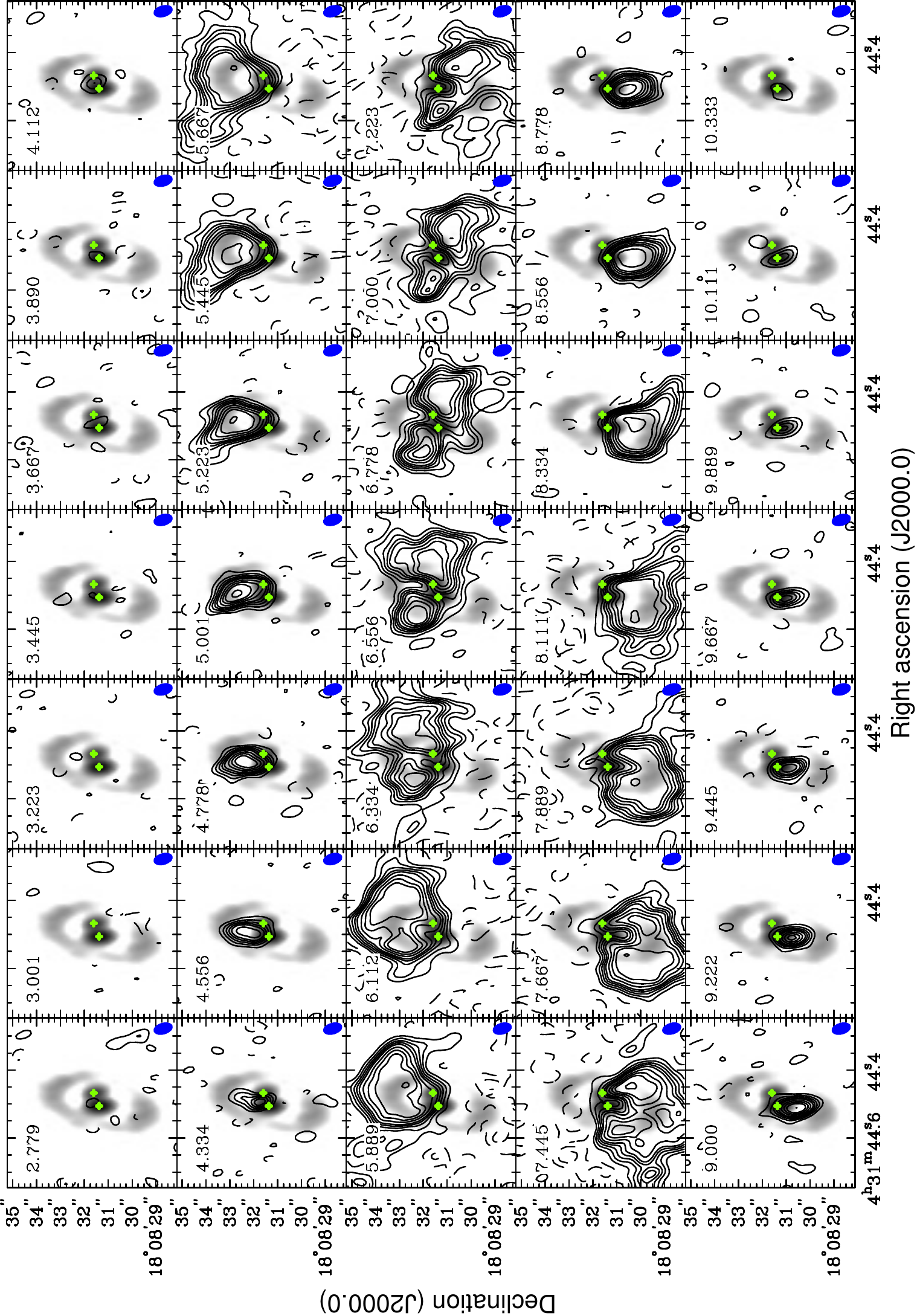}
\caption{Theoretically-predicted velocity channel maps of the C$^{18}$O ($J$=3-2) emission
(contours) superposed on the theoretically-predicted 0.9-mm dust-continuum emission (grey) in L1551 NE,
calculated from our hydrodynamic
and radiative transfer calculations and the ALMA observing simulation. Contour levels
and symbols are the same as those in Figure \ref{c18och}.
\label{modelch}}
\end{figure}

From the model distribution of matter in the circumstellar and circumbinary disks, we performed
radiative transfer calculation so as to permit a direct comparison between the theoretical images
and those obtained in our ALMA observation. 
To derive the emission distributions of the model images,
we simply integrated each specific intensity in each cell along the light of sight using
the analytic formula of the radiative transfer and assuming the
local thermodynamic equilibrium (LTE) condition ($i.e.$, $T_{ex}$ = $T_{k}$).
For the purpose of producing more realistic theoretical images,
we assumed a power-law temperature distribution of $T \propto r^{-0.2}$,
whereas for simplicity we assumed the gas to be isothermal in the hydrodynamic simulation
(Detailed discussion on the assumed temperature is given in section 4.1.).
The assumed dust opacity and C$^{18}$O abundance are listed in Table 2.
Finally, the theoretical images are passed through the
CASA task ``simobserve'' to create simulated visibility data,
and the simulated visibility data are Fourier-transformed and
de-convolved to create the simulated images, shown in
Figures \ref{mom1s}, \ref{c18opvs}, \ref{contall}, and \ref{modelch}.
In this step, we replicated the same antenna
configuration, observing time and hour angle convergence, data
flagging, thermal noise, and imaging parameters as the actual observation and data processing.

\clearpage

\clearpage

\clearpage

\clearpage

\clearpage

\clearpage

\clearpage

\clearpage

\clearpage

\clearpage

\clearpage

\clearpage
\begin{deluxetable}{lc}
\tablecaption{Parameters for the ALMA Observation of L1551 NE \label{tbl-1}}
\tabletypesize{\scriptsize}
\tablewidth{0pt}
\tablehead{\colhead{Parameter} &\colhead{Value}}
\startdata
Observing date & 2012 Nov. 18\\
Number of antennas & 25\\
Right ascension (J2000.0) & 04$^{h}$31$^{m}$44$\fs$5 \\
Declination (J2000.0) & 18$\degr$08$\arcmin$31$\farcs$67 \\
Central Frequency (continuum) & 335.85 GHz\\
Primary beam HPBW & $\sim$18$\arcsec$ \\
Synthesized beam HPBW (C$^{18}$O) & 0$\farcs$82$\times$0$\farcs$49 (P.A. = 16$\degr$)  \\
Synthesized beam HPBW (continuum; $>$80 $k\lambda$) & 0$\farcs$72$\times$0$\farcs$36 (P.A. = 9$\degr$) \\
Baseline coverage & 14.9 - 368.0 m \\
Conversion Factor (C$^{18}$O) & 1 (Jy beam$^{-1}$) = 27.9 (K)\\
Frequency resolution & 244.14 kHz $\sim$0.22 km s$^{-1}$\\
Bandwidth (continuum) & 1.79 GHz \\
Flux calibrator &Callisto\\
Gain calibrator &J0510+180 \\
Flux (J0510+180) &1.9 Jy \\
Passband calibrator    &J0423-0120  \\
System temperature &$\sim$150 - 400 K \\
rms noise level (continuum)& 2.6 mJy beam$^{-1}$\\
rms noise level (C$^{18}$O)& 15.8 mJy beam$^{-1}$\\
\enddata
\end{deluxetable}

\clearpage
\begin{deluxetable}{ll}
\tabletypesize{\scriptsize}
\tablecaption{Parameters for the Theoretical Model of L1551 NE \label{tbl-2}}
\tablewidth{0pt}
\tablehead{\colhead{Parameter} &\colhead{Value}}
\startdata
Computer & ATERUI in NAOJ CfCA\\
Simulation Code & 3D AMR Code (SFUMARTO; Matsumoto 2007)\\
Execution Time & 35 hours (512 Cores) \\
Simulation Box & (1740 AU)$^3$\\
Highest Resolution &0.85 AU\\
Radius of Sink Particles &3.4 AU\\
Boundary Radius &1740 AU\\
Image Pixel Size & 5 AU \\
Binary Separation$\tablenotemark{a}$         & 145 AU \\
Centrifugal Radius of Gas$\tablenotemark{a}$ & 300 AU\\
Disk Position Angle$\tablenotemark{a}$ & 167$^{\circ}$\\
Disk Inclination Angle$\tablenotemark{a}$      & -62$^{\circ}$  \\
Total Binary Mass$\tablenotemark{a}$                 &  0.8 $M_{\odot}$    \\
Binary Mass Ratio$\tablenotemark{a}$ & 0.19   \\
Gas Number Density at the Boundary & 1.5 $\times$ 10$^{5}$ cm$^{-3}$\\
Mean Molecular Weight & 2.3 \\
Temperature Profile$\tablenotemark{b}$ &
$T(r) = \max\left[ 23~{\rm K} \left(\frac{r_A}{300~{\rm AU}}\right)^{-0.2},
19~{\rm K} \left(\frac{r_B}{300~{\rm AU}}\right)^{-0.2} \right]$
\\
Dust Opacity$\tablenotemark{c}$   &
$\kappa_{0.9~mm} = 0.053~\mathrm{cm}^{2}~\mathrm{g}^{-1}$ \\
C$^{18}$O Abundance$\tablenotemark{d}$ & $1.7 \times 10^{-7}$ \\
\enddata
\tablenotetext{a}{Derived from the Keplerian-disk model fitting to the C$^{18}$O data
taken with the SMA (Takakuwa et al. 2012).}
\tablenotetext{b}{Radii $r_A$ and $r_B$ indicate the distance from Source A and that from
Source B, respectively.}
\tablenotetext{c}{See Section 3.1.}
\tablenotetext{d}{Crapsi et al. (2004).}
\end{deluxetable}

\end{document}